\documentclass[
  journal=pasa,
  manuscript=research-paper, 
  year=2022,
  volume=xx,
]{cup-journal}

\usepackage{microtype,siunitx,booktabs,graphicx,color,rotating,amsmath,amssymb,tabularx}
\title{WALLABY Pre-Pilot Survey: Radio Continuum Properties of the Eridanus Supergroup}

\author{J. A. Grundy}
\affiliation{International Centre for Radio Astronomy Research, Curtin University, Bentley, WA6102, Australia}
\alsoaffiliation{CSIRO Space and Astronomy, PO Box 1130, Bentley, WA6102, Australia}
\email[J. A. Grundy]{joe.grundy@postgrad.curtin.edu.au}

\author{O. I. Wong}
\affiliation{CSIRO Space and Astronomy, PO Box 1130, Bentley, WA6102, Australia}
\alsoaffiliation{International Centre for Radio Astronomy Research, University of Western Australia, 35 Stirling Hwy, Crawley, WA6009, Australia}
\alsoaffiliation{ARC Centre of Excellence for Astrophysics in Three Dimensions (ASTRO 3D), Australia}

\author{K. Lee-Waddell}
\affiliation{International Centre for Radio Astronomy Research, University of Western Australia, 35 Stirling Hwy, Crawley, WA6009, Australia}
\alsoaffiliation{CSIRO Space and Astronomy, PO Box 1130, Bentley, WA6102, Australia}
\alsoaffiliation{International Centre for Radio Astronomy Research, Curtin University, Bentley, WA6102, Australia}

\author{N. Seymour}
\affiliation{International Centre for Radio Astronomy Research, Curtin University, Bentley, WA6102, Australia}

\author{B.-Q. For}
\affiliation{International Centre for Radio Astronomy Research, University of Western Australia, 35 Stirling Hwy, Crawley, WA6009, Australia}
\alsoaffiliation{ARC Centre of Excellence for Astrophysics in Three Dimensions (ASTRO 3D), Australia}

\author{C. Murugeshan}
\affiliation{CSIRO Space and Astronomy, PO Box 1130, Bentley, WA6102, Australia}
\alsoaffiliation{ARC Centre of Excellence for Astrophysics in Three Dimensions (ASTRO 3D), Australia}

\author{B. S. Koribalski}
\affiliation{CSIRO Astronomy and Space Science, Australia Telescope National Facility, P.O. Box 76, NSW1710, Australia}
\alsoaffiliation{School of Science, Western Sydney University, Locked Bag 1797, Penrith, NSW2751, Australia}

\author{J. P. Madrid}
\affiliation{The University of Texas Rio Grande Valley, One West University Blvd, Brownsville, TX78520 USA}

\author{J. Rhee}
\affiliation{International Centre for Radio Astronomy Research, University of Western Australia, 35 Stirling Hwy, Crawley, WA6009, Australia}
\alsoaffiliation{ARC Centre of Excellence for Astrophysics in Three Dimensions (ASTRO 3D), Australia}

\author{T. Westmeier}
\affiliation{International Centre for Radio Astronomy Research, University of Western Australia, 35 Stirling Hwy, Crawley, WA6009, Australia}
\alsoaffiliation{ARC Centre of Excellence for Astrophysics in Three Dimensions (ASTRO 3D), Australia}

\doi{xx} 

\received {20 Sep 2022}
\revised  {24 Jan 2023}
\accepted {23 Feb 2023}
\published{dd Mmm YYYY}


\begin{document}

\begin{abstract}
We present the highest resolution and sensitivity $\sim1.4\,$GHz continuum observations of the Eridanus supergroup obtained as a part of the Widefield ASKAP \emph{L}-band Legacy All-sky Blind surveY (WALLABY) pre-pilot observations using the Australian Square Kilometer Array Pathfinder (ASKAP). We detect 9461 sources at 1.37\,GHz down to a flux density limit of $\sim0.1$\,mJy at $6.1''\times 7.9''$ resolution with a mean root-mean-square (RMS) of 0.05 mJy/beam. We find that the flux scale is accurate to within 5\,\% (compared to NVSS at 1.4\,GHz). We then determine the global properties of eight Eridanus supergroup members, which are detected in both radio continuum and neutral hydrogen (HI) emission, and find that the radio-derived star formation rates (SFRs) agree well with previous literature. Using our global and resolved radio continuum properties of the nearby Eridanus galaxies, we measure and extend the infrared-radio correlation (IRRC) to lower stellar masses and inferred star formation rates than before. We find the resolved IRRC to be useful for: 1) discriminating between AGN and star-forming galaxies (SFGs); 2) identifying background radio sources; and 3) tracing the effects of group environment pre-processing in NGC 1385.  We find evidence for tidal interactions and ram-pressure stripping in the HI, resolved spectral index and IRRC morphologies of NGC 1385. There appears to be a spatial coincidence (in projection) of double-lobed radio jets with the central HI hole of NGC 1367.  The destruction of polycyclic aromatic hydrocarbons (PAHs) by merger-induced shocks may be driving the observed WISE W3 deficit observed in NGC 1359. Our results suggest that resolved radio continuum and IRRC studies are excellent tracers of the physical processes that drive galaxy evolution and will be possible on larger sample of sources with upcoming ASKAP radio continuum surveys.
\end{abstract}

\section{Introduction}

Star formation (SF) is one of the dominant physical processes in the formation and evolution of galaxies. The majority of the star formation rate (SFR) indicators in galaxies are either indirect indicators of young stars and/or affected by dust extinction due to being in the ultraviolet and the optical bands of the electromagnetic spectrum. Radio continuum emission is optically thin at GHz frequencies and has been shown to be a dust unbiased tracer of recent SF within galaxies \citep{Davies2017}. Synchrotron emission is produced primarily by the acceleration of cosmic ray (CR) electrons by shocks in a galaxy's magnetic field after young massive stars ($\geq8$\,M$_{\odot}$) undergo core-collapse supernova. These CR electrons undergo a number of different cooling processes as they propagate throughout the galaxy including inverse Compton (IC), free-free and ionization losses \citep{Murphy2009}. The ultraviolet (UV) light from young fairly massive stars ($\geq5$\,M$_{\odot}$) also ionises the surrounding hydrogen gas cloud producing thermal (free-free) radio emission and heating the nearby dust which re-radiates in the infrared (IR) regime. Synchrotron emission, which composes a majority of the radio emission at low frequencies \citep[$\lesssim$30\,GHz;][]{Condon1992, Yun2001, Bell2003}, is dominated by active galactic nuclei (AGN) for most bright radio sources. However, as the sensitivity and resolution of large scale radio surveys has improved, we find that they are uncovering a large population of low-redshift SFGs. In order to utilise the strengths of radio continuum emission as a SF tracer it has been calibrated against previously established SF tracers such as $H\alpha$, UV and IR emission down to 1\,mJy using the Faint Images of the Radio Sky at Twenty centimeters (FIRST) survey in the northern hemisphere and Galaxy and Mass Assembly (GAMA) fields \citep{Davies2017, Molnar2021}. Radio-SFR relations are found have a tight spread comparable to widely used SF tracers \citep[$\sim$0.15 dex;][]{Tabatabaei2017} and a (mostly) linear correlation to the total IR emission \citep[$\sim$0.16 dex;][]{Molnar2021}. 

The correlation between the far or total IR radiation and the radio continuum emission is known as the "infrared-radio correlation" (IRRC) and has been studied extensively \citep[][]{Helou1985, Condon1992, Yun2001, Bell2003, Delhaize2017, Molnar2021, Delvecchio2021} on the low-z IRRC; however, the physical processes that govern this relationship are still poorly understood. The IRRC, which is usually parameterised by the $q$ value (Eq.~\ref{eq:qratio}; the ratio of IR to radio luminosities), has been examined against the variation of other galaxy properties such as stellar mass \citep{Magnelli2015, Delvecchio2021} and galaxy type \citep{Moric2010, Nyland2017} and appears to hold over at least three orders of magnitude in both radio and IR luminosity \citep{Helou1985, Condon1992, Delvecchio2021, Molnar2021} even in merging galaxies \citep{Condon2002, Murphy2013}. Another area of interest is the evolution of the IRRC with redshift, which has been a hotly debated topic \citep{Jarvis2010, Sargent2010, Magnelli2015, Delhaize2017} with recent studies finding a slight decrease in the $q$ value as redshift increases \cite{Delhaize2017, Molnar2021, Delvecchio2021}. This tight, linear correlation is upheld by a number of physical processes and has been investigated in terms of both calorimetric and non-calorimetric models. Calorimetric models assume that the galaxies are optically thick to UV such that their UV emission is fully re-radiated in the IR and that the CR electrons radiate away their total energy as synchrotron emission before escaping the galaxy \citep{Volk1989} however this implies that the IRRC is likely to break down towards lower mass galaxies \citep{Bourne2012}. Whilst, non-Calorimetric or optically thin models \citep{Helou1993, Bell2003, Lacki2010, Vollmer2022} argue a sort of ``conspiracy'' between physical mechanisms which acts to maintain the perhaps unexpected linearity of the IRRC. However, studies by \citet{Murphy2009} have also found the $q$ value to be systematically lower in cluster environments in comparison to field galaxies suggesting that environmental effects may affect the physical processes which give rise to the IRRC. Thus understanding the physical mechanisms which cause these discrepancies between observations and model predictions is crucial to our use of the IRRC and radio emission as a SFR tracer.

While there are many corroborating studies on the existence of the IRRC in SFGs there has been limited research on how the IRRC varies \textit{within} galaxies, primarily due to the limited resolution and sensitivity of past IR and radio surveys \cite{Murphy2008, Murphy2009, Tabatabaei2013, Vollmer2020}. We expect variations in interstellar medium (ISM) properties and the effects of environmental interactions to lead to observed variations of the IRRC within individual galaxies \cite{Murphy2009, Tabatabaei2013, Vollmer2020}. As synchrotron emission is dependent on both the energy density of CR electrons and magnetic field strength, physical mechanisms which modify these properties will impact the IRRC. Furthermore the significant difference in mean free path of CR electrons (1-2 kpc) relative to that dust heating photons ($\sim100$\,pc) suggests that variations in the IRRC can be expected \cite{Bicay1990}. \citet[][]{Hughes2006} finds that the synchrotron haloes around individual star-forming regions extend further than their corresponding FIR emission within the Large Magellanic Cloud. \citet[][]{Murphy2009, Vollmer2020} show that Virgo cluster galaxies experiencing intra-cluster medium (ICM)-ISM interactions often display radio-deficit regions at the interacting edge of the galaxy and synchrotron tails caused by the re-acceleration of the CR electrons in the ICM driven shocks. A correlation between SFR density and magnetic field strength is observed in \citet{Tabatabaei2013} and \citet{Heesen2022a, Heesen2022b} suggesting that environmental effects which can enhance or quench SF in different parts of galaxies will also cause discrepancies in the IRRC within galaxies. 

\subsection{Eridanus Supergroup}

The Widefield ASKAP {\it L}-band Legacy All-sky Blind surveY (WALLABY; \citet{Koribalski2020}) is conducted with the Australian Square Kilometre Array Pathfinder (ASKAP; \citet{Hotan2021}) and will utilise the telescope's large field of view to image hundreds of thousands of HI galaxies out to a redshift of $z \sim 0.26$ \citet{Koribalski2020}. The WALLABY team has performed a pre-pilot survey of the Eridanus region that contains three distinct galaxy groups namely the NGC 1407, NGC 1332 and Eridanus (or NGC 1395) groups, which form a `supergroup' or a group of groups that may eventually merge to form a cluster \citep[][]{Willmer1989, Omar2005, Brough2006, For2021}. \citet{Brough2006} find the NGC 1407 group to be the most evolved and contains a larger fraction of early-type galaxies compared to the NGC 1332 and Eridanus groups. Studies of the disturbed HI morphologies from recent WALLABY observations of the Eridanus galaxies suggest that these galaxies are currently undergoing different stages of tidal interactions and ram-pressure stripping \citep[][]{For2021, Murugeshan2021, Wangs2022}.

An additional product of the Eridanus pre-pilot HI observations are high-resolution radio continuum images centred at $\sim1.37$\,GHz. Using these observations we aim to determine if the environmental effects and pre-processing, which have been shown to be affecting the HI gas and impacting the SF properties of galaxies in the Eridanus supergroup, manifest in the radio continuum emission. Utilising NASA's {\it Wide-field Infrared Survey Explorer} \citep[{\it WISE};][]{Wright2010} observations we also examine the effect that environmental pre-processing has on the IRRC both globally and within individual galaxies. Full WALLABY as well as other upcoming radio continuum surveys such as the Evolutionary Map of the Universe (EMU; \citet{Norris2011}) will allow a more thorough investigation of the radio continuum and hence SF properties within individual galaxies. These future observations will also allow us to examine the resolved IRRC of individual galaxies and provide the means to understand the physical processes that give rise to this relationship and what may cause it to break down.

This paper complements the other WALLABY Pre-Pilot Survey papers by exploring the impact of the environmental interactions on the SF and ISM properties probed using the IRRC and improving our understanding of the physical origin of this correlation. In section 2, we provide details about the observations, data processing, and source extraction for the WALLABY continuum (WC) catalogue and the Eridanus supergroup members. Section 3 describes the ancillary data used. Section 4 presents our results including a quality assessment of the WC catalogue. Section 4 also presents the global and resolved SFR, IRRC, and spectral index properties of eight Eridanus supergroup galaxies. In section 5, we discuss our results including possible interpretations and limitations. Section 6 concludes our findings. Throughout the paper, we adopt a $\Lambda$ cold dark matter cosmology model ($\Lambda$CDM) with $\Omega_{\rm M}$ = 0.27, $\Omega_{\rm K}$ = 0, $\Omega_{\Lambda}$ = 0.73 and $H_{0}$ = 73\,${\rm km s^{-1} Mpc^{-1}}$. Following \citet{For2021} we assume a distance to the Eridanus supergroup of 21\,Mpc.

\section{WALLABY Data}

\subsection{Observations and Data Processing}

The Murchison Radio astronomy Observatory (MRO) -- located in the remote outback of Western Australia -- hosts ASKAP, a radio interferometer consisting of 36 12-m antennas. Each antenna is equipped with a second generation MK II phased array feed from which 36 dual-polarisation beams are formed. These beams are designed to provide an instantaneous field of view of $5.5^{\circ}\times5.5^{\circ}$ (at $\sim$1.3 GHz) with a bandwidth of 288\,MHz \citep{McConnell2016, Hotan2021}. WALLABY carried out a pre-pilot survey in March 2019 targeting the Eridanus galaxy supergroup field (hereafter the Eridanus field) which utilised the full array for two interleaving footprints (footprints A and B) to obtain uniform sensitivity.

\begin{figure*}[hbt!]
	\includegraphics[width=\textwidth]{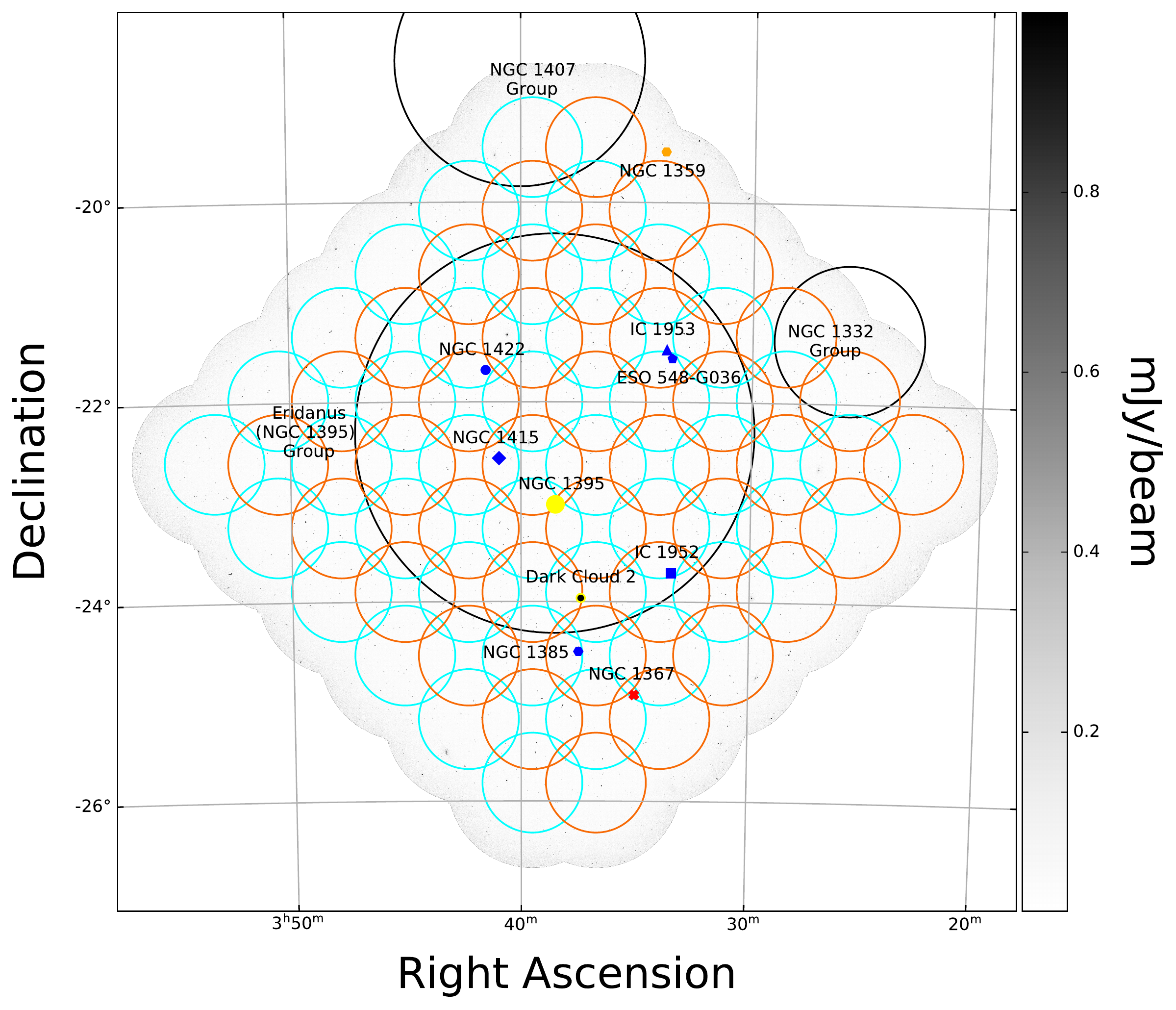}
    \centering
    \caption{WALLABY continuum image of the entire Eridanus field including locations of the Eridanus (NGC 1395) group members marked in blue (or red for the AGN) with the group central galaxy NGC 1395 marked in yellow. Dark cloud 2, an HI cloud with no detected stellar component \citep{Wong2021} is also marked in black. The detected NGC 1407 group member NGC 1359 is marked in orange. The cyan and orange circles represent footprints A and B respectively for the ASKAP observations listed in Table~\ref{tab:Obstab} and each beam has a FWHM of $1^{\circ}$. The black circles mark the maximum radial extent of the groups belonging to the Eridanus supergroup as identified in \citet{Brough2006}.}
    \label{fig:WALfield}
\end{figure*}

Each footprint has a $6 \times 6$ beam pattern as shown in Fig.~\ref{fig:WALfield}. The Eridanus field central coordinates for footprints A and B are given in Table~\ref{tab:Obstab} along with other observing parameters. The observations for footprint A were carried out during the day and those for footprint B were obtained mostly at night. The primary calibrator PKS 1934-638 was observed for 2-3 hours between the two sets of science observations. Each observation has a scheduling block identification number (SBID) that can be used to access the corresponding public data set in the CSIRO ASKAP Science Data Archive\footnote{https://research.csiro.au/casda/} (CASDA). 

The observations were processed using the ASKAPSOFT pipeline version 0.24.7\footnote{https://www.atnf.csiro.au/computing/software/askapsoft/sdp/docs/\newline current/general/releaseNotes.html\#october-2019} \citep{Whiting2020} with WALLABY-specific processing parameters. Only the upper half of the 288 MHz bandwidth (1295.5-1439.5\,MHz, centred at 1367.5\,MHz) was processed due to satellite radio frequency interference (RFI) in the lower half. Footprints A and B (Fig.~\ref{fig:WALfield}) were processed separately. For a detailed description of the processing steps refer to \citet{For2021}. The calibrated visibilities for all baselines were used for the radio continuum, whereas imaging for the WALLABY HI cubes only used baselines up to two kilometres. The continuum visibilities were CLEANed, after applying a Wiener filter, with robustness value of -0.5 and no applied taper. A single loop of automated self-calibration in included in the ASKAPSOFT continuum pipeline. After imaging, the two individual footprints were linearly mosaicked to produce the final images and data cubes. These images are not convolved to a common resolution which is part of the current pipeline for processing of radio continuum images with ASKAP and has been shown to produce more accurate flux density measurements \cite{Norris2021, Hale2021}.

These continuum images produced by WALLABY are the highest resolution and sensitivity $\sim 1.4\,$GHz observations available for the Eridanus field, boasting an angular resolution of 7.9$''$ $\times$ 6.1$''$ with integrated flux densities between 0.14 and 831.6\,mJy, a dynamic range of $\sim1.2\times10^4$, and a median RMS noise value of 50\,$\mu$Jy/beam.

\subsection{Source Extraction}

Source extraction and flux measurements for WALLABY radio continuum sources were performed using the PROFOUND function within {\tt ProFound} \citep{Robotham2018}, which has been shown to be an excellent pixel-based source extraction tool for high resolution radio surveys that exhibit both complex and simple sources \citep[][]{Hale2019}. The full details of how {\tt ProFound} works are given in \citet{Robotham2018} and we refer the reader to \citet[][section 2.1]{Hale2019} for a brief description of its application to radio data. Parameters were left at default values outside of those shown in Table~A.\ref{tab:inptab}, which were selected in order to identify a large number of initial seeds that are then dilated to allow the merging of extended/diffuse segments (islands which encapsulate the flux of a radio source). Modification of these parameters largely affects the number of faint sources detected (box, skycut, pixcut) or whether extended/diffuse sources are correctly combined into one segment (tolerance, reltol, ext). Final values were chosen based on test fields from different regions of the Eridanus field containing a number of different radio features and noise levels to achieve a balance between completeness and reliability.

\begin{figure}[hbt!]
	\includegraphics[width=\columnwidth]{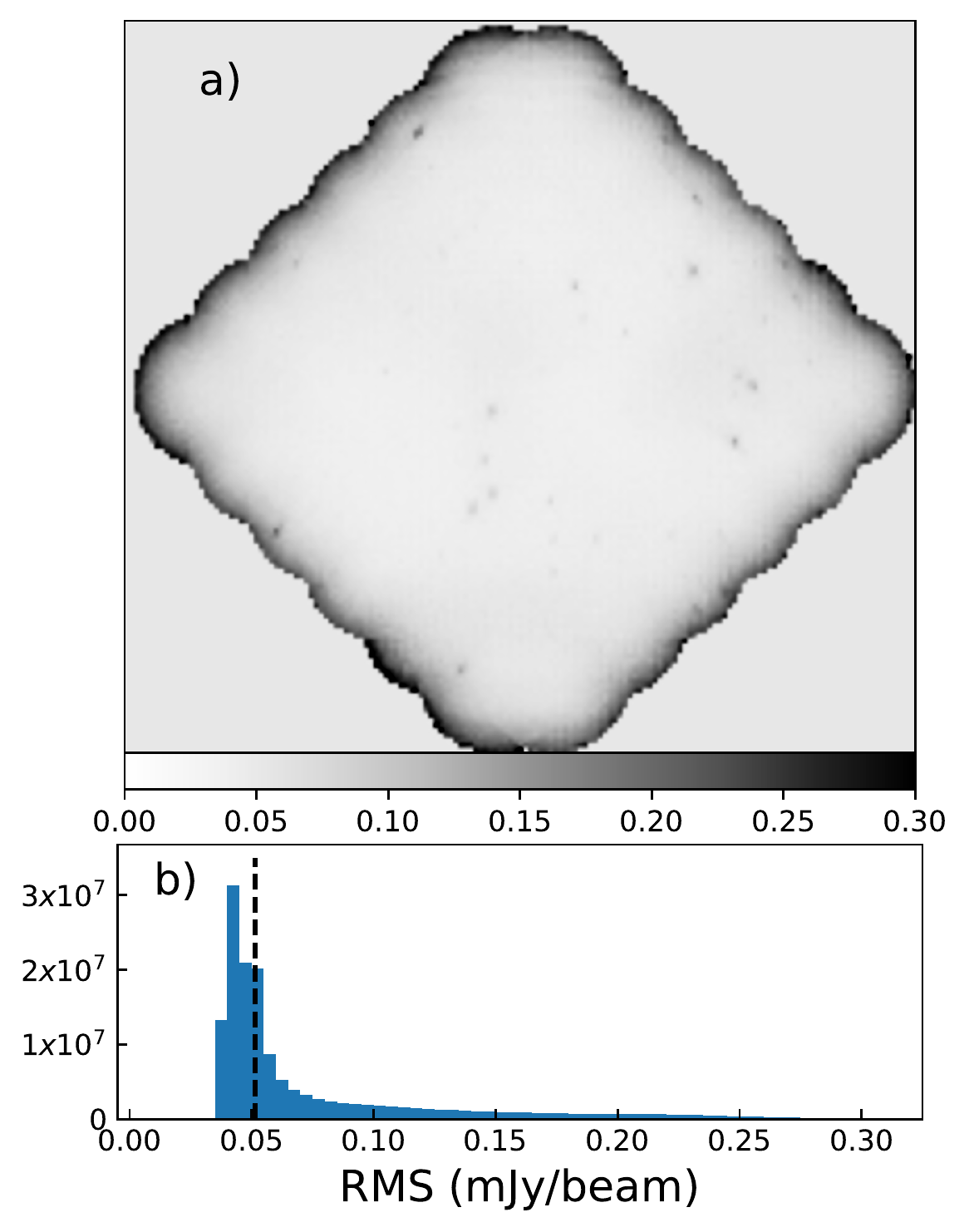}
    \caption{Top: map of the RMS noise in the Eridanus field (with the colour bar in units of mJy/beam). Bottom: histogram of RMS values with a median value of 0.05\,mJy/beam shown by the dashed line.}
    \label{fig:RMSmap}
\end{figure}

Initial source extraction from {\tt ProFound} finds 9948 components that are then cut at a 5$\sigma$ threshold. This 5$\sigma$ threshold is defined by the final peak flux value divided by the mean segment RMS noise for each component since the initial skycut of 3$\sigma$ is performed before accurate backgrounds and RMS maps are measured. We note that significant sidelobes from the dirty beam occur around bright components ($S_{\rm Peak}\gtrsim50\,$mJy) in a spiral residual pattern which are often detected as individual components by {\tt ProFound}. In order to remove these sidelobes from the component list we remove segments within 180$''$ of bright components that have SNR<10$\sigma$. Upon visual inspection, this criteria correctly removes sidelobes around all but one bright component that exists in the outer edge of the observed field, which has no footprint overlap. For this component, the three significant segments that are consistent with the spiral residual pattern were manually removed from the component catalogue. Real faint sources that are within this residual pattern may also be removed using this criteria. Visual inspection or machine learning methods are often required to correctly identify complex radio sources which have multiple associated components. No attempt was made to classify components as radio objects (doubles/triples) as this is beyond the scope of this work. These issues will be better addressed in the WALLABY pilot survey and WALLABY full survey as better primary beam models from holography measurements are used \citep{Norris2021, Hale2021}. Overall the final WALLABY continuum (WC) catalogue contains 9461 components above a $5\sigma$ threshold with a resolution of 7.9$''$ $\times$ 6.1$''$.  In the top panel of Fig.~\ref{fig:RMSmap}, we show the uniformity of the noise across the field with localised increases due to low-level calibration errors around bright sources and at the edges of the field where the beam sensitivities drop off. This noise at the edge of the field causes the extended RMS tail as seen in the bottom panel of Fig.~\ref{fig:RMSmap}.

\subsection{Identifying Unresolved Sources}
\label{sec:unres}

\begin{figure}[hbt!]
	\includegraphics[width=\columnwidth]{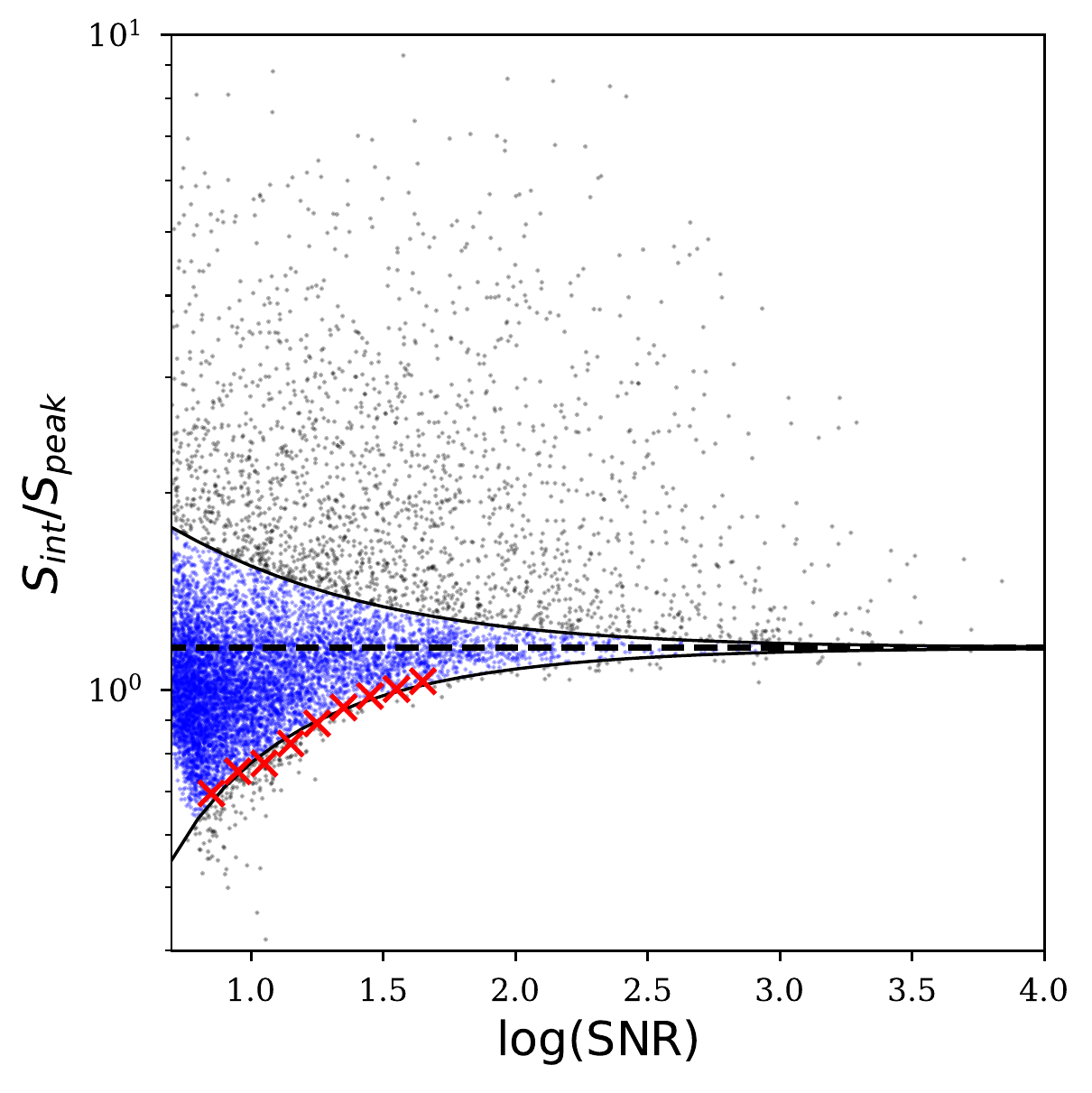}
    \centering
    \caption{The ratio of integrated to peak flux as a function of SNR for sources with SNR > 5 and the envelope used to define unresolved sources. Blue points indicate sources which are unresolved by our criteria and grey points indicate resolved sources as defined by the envelope. The red crosses indicates the $S_{\rm int}/S_{\rm peak}$ values in which 95$\%$ of the sources below $S_{\rm int}/S_{\rm peak} = 1.16$ are included within the envelope for bins of width 0.1 dex and including greater than 50 sources. The black dashed line indicates $S_{\rm int}/S_{\rm peak} = 1.16$.}
    \label{fig:envelopefig}
\end{figure}

We select unresolved sources for comparisons between surveys using the method employed for the Rapid ASKAP Continuum Survey \citep[RACS;][]{McConnell2020, Hale2021} which follows from \citet{Bondi2008}, \citet{Smolcic2017} and \citet{Shimwell2019} and involves defining an envelope used to distinguish between resolved and unresolved sources. We construct this envelope by selecting sources with SNR $\geq 5$ where SNR is defined by the peak flux of a segment divided by the estimated mean background RMS noise of that segment. We then measure the ratio of integrated flux ($S_{\rm int}$ to peak flux ($S_{\rm peak}$) as a function of SNR. For unresolved sources the ratio of integrated to peak flux ($S_{\rm int}/S_{\rm peak}$) should equal one by construction if the synthesised beam size is an accurate representation of image resolution. Typically we see a scatter around unity especially at low SNR where noise has a greater impact on flux measurements. We find that for our data $S_{\rm int}/S_{\rm peak}$ tends towards 1.16 (as measured by the highest SNR unresolved point source) likely due to unmodelled source smearing possibly due to tracking errors or inappropriate primary beam modelling. The discrepancy of this value does not impact our analysis and construction of the envelope. Following \citet{Hale2021} we expect unresolved sources to have values of $S_{\rm int}/S_{\rm peak}$ as a function of SNR which lie between the envelopes as described by: 
\begin{equation}
    \frac{S_{\rm int}}{S_{\rm peak\pm}} = 1.16 \pm A \times {\rm SNR}^{-B}.
	\label{eq:envelope}
\end{equation}
We determine values of A and B from the lower envelope shown in Fig. \ref{fig:envelopefig}, $S_{\rm int}/S_{\rm peak-}$, and declare sources with $S_{\rm int}/S_{\rm peak}$ > $S_{\rm int}/S_{\rm peak+}$ to be resolved as they will have elevated $S_{\rm int}/S_{\rm peak}$ values. We generate this fit using equally spaced bins in log(SNR) (excluding the lowest SNR bin) which each contain at least 50 sources. Then the $S_{\rm int}/S_{\rm peak}$ ratio that contains 95$\%$ of the sources with $S_{\rm int}/S_{\rm peak}$ < 1.16 is determined. These points are then fit to the lower envelope of Eq.~\ref{eq:envelope} which yields $1.16 - 1.78 \times \mbox{SNR}^{-0.66}$. The lower envelope is reflected around $S_{\rm int}/S_{\rm peak}$ = 1.16 to create the upper envelope. Sources which lie between these two envelopes are defined as unresolved and are shown in Fig.~\ref{fig:envelopefig}. We find that $\sim$76.5\% of WC sources are unresolved at a resolution of $7.9'' \times 6.1''$ and hence should also be unresolved in the National Radio Astronomy Observatory (NRAO) Very Large Array (VLA) Sky Survey \citep[NVSS;][]{Condon1998} and RACS.

Sources are then selected from WC and matched to comparison surveys using the same four criteria as described in \citet{Hale2021}:
\begin{enumerate}
    \item Sources are isolated to within an angular separation of twice the FWHM of the comparison survey with poorer angular resolution i.e. $50''$ and $90''$ angular separations for comparisons with RACS and NVSS respectively.
    \item Sources have a peak SNR $\geq 10$ in WC.
    \item Sources are unresolved as described by the envelope criterion above.
    \item Sources match positions within an angular separation of $10''$, which allows for variation in the positions of measured sources given the different angular resolutions of the comparison surveys.
\end{enumerate}
After which, we find 1418 matches between WC and RACS, and 384 matches common to all three catalogues.

\subsection{Eridanus Supergroup}

In order to examine whether the environmental effects within supergroups are reflected in the radio continuum emission, we select galaxies that are known to be interacting with their local environment. The WALLABY HI observations in the Eridanus supergroup find that almost all HI-detected group member galaxies down to an HI mass of $10^{7.7}\,M_{\odot}$ have disturbed HI morphologies \citep{For2021}. Most group members are observed to be HI deficient compared to their stellar masses and integrated disk stability suggesting that gas has been removed from the galaxies due to tidal interactions, mergers and ram pressure stripping \citep[][]{For2021, Murugeshan2021, Wangs2022}. We look to investigate the radio continuum emission and derived physical properties to determine whether we can measure the impact that environment has on these SFGs.

\subsubsection{Sample Selection}

The catalogue of 43 Eridanus supergroup HI sources detected in the WALLABY pre-pilot survey \citep{For2021} are positionally matched -- within 10$''$ -- to sources in the WC catalogue. We detect eight galaxies with corresponding radio continuum emission which are shown in Fig.~\ref{fig:Murphy1}. There was also one background star-forming galaxy with radio continuum emission that was incorrectly cross-matched to LEDA 13743 and not included in our sample. The measured integrated flux densities at multiple wavelengths are presented for each galaxy along with their position and morphology in Table~\ref{tab:WCmeasured}. NVSS \citep{Condon1998}, RACS \citep{Hale2021} and, GMRT \citep{Omar2005} integrated flux values are obtained for the eight WC-selected galaxies whilst the WC flux densities are measured using {\tt ProFound}. To ensure we encapsulate all the flux from the observed isolated radio continuum components in the star-forming disks of IC 1952, IC 1953 and NGC 1359 {\it WISE} W1 segment maps are passed to {\tt ProFound}. The remaining Eridanus supergroup members use the measured WC catalogue flux densities as their less complex radio morphology allows correct assignment and measurement of their total flux density. 

\begin{table*}[hbt!]
\centering
\begin{threeparttable}
\caption{Radio continuum properties of Eridanus supergroup members.}\label{tab:WCmeasured}
\begin{tabular}{c|c|c|c|c|c|c|c|c}
    \toprule
	Designation & Common ID & R.A & Dec & Morphology & $S_{\rm int}^{\rm WC}$ & $S_{\rm int}^{\rm NVSS}$ & $S_{\rm int}^{\rm RACS}$ & $S_{\rm int}^{\rm O05}$ \\
	&  & ($^{\circ}$) & ($^{\circ}$) &  & (mJy) & (mJy) & (mJy) & (mJy) \\
    \midrule
	WC J033326-234229 & IC 1952 & 53.359759 & -23.708036 & SB(s)bc? &  $1.0\pm0.3^{a,b}$ & $4.5\pm0.6$ & $ 8\pm 1$ & $ 7\pm 1$ \\
	WC J033328-213353 & ESO 548-G036 & 53.365284 & -21.565734 & Sc &  $7.9\pm1.1$ & $7.6\pm0.5$ & $13\pm 2$ & $ 9\pm 1$ \\
	WC J033342-212843 & IC 1953 & 53.424245 & -21.478690 & SB(rs)d &  $6.9\pm1.4^{b}$ & $13.0\pm2.2$ & $ 5\pm 1$ & $ 9\pm 2$ \\
	WC J033348-192935 & NGC 1359 & 53.448790 & -19.492060 & SB(s)m? & $18.9\pm3.6^{b}$ & $31.1\pm2.4$ & $40\pm 3$ & $32\pm 3$ \\
	WC J033501-245558 & NGC 1367 & 53.754798 & -24.932908 & SAB(rs)a &  $13.2\pm1.8$ & $15.4\pm1.8$ & $25\pm 3$ & $20\pm 2$ \\
	WC J033729-243001 & NGC 1385 & 54.369089 & -24.500355 & SB(s)cd &  $142.2\pm18.7$ & $178.5\pm6.1$ & $246\pm18$ & $180\pm20$ \\
	WC J034057-223353 & NGC 1415 & 55.236565 & -22.564758 & (R)SAB0/a(s) &  $23.4\pm3.1$ & $25.3\pm1.2$ & $36\pm 3$ & $27\pm 3$ \\
	WC J034132-214049 & NGC 1422 & 55.382081 & -21.680283 & SBab &  $0.9\pm0.2$ &  &  &  \\
    \bottomrule
\end{tabular}
\begin{tablenotes}[para]
	\item[]Note: Column (1): WC designation. Column (2): Galaxy common ID. Column (3): R.A of the WC flux weighted centre. Column (4): Declination of the WC flux weighted centre. Column (5): Optical morphology from NED. Column (6): 1.37\,GHz WC total integrated flux density. Column (7): 1.4\,GHz NVSS total integrated flux density. Column (8): 0.88\,GHz RACS total integrated flux density. Column (9): 1.4\,GHz \citet{Omar2005} total integrated flux density. 
    \item[$^{a}$]The background contribution is removed. 
    \item[$^{b}$]We measure WC integrated flux density using an aperture which encapsulates the disk and core continuum emission.
\end{tablenotes}
\end{threeparttable}
\end{table*}

\section{Ancillary Data}

\begin{table*}[hbt!]
\centering
\begin{threeparttable}
\caption{Basic survey information.}\label{tab:surveytab}
\begin{tabular}{cccccc}
    \toprule
	Survey & $\nu (\lambda)$ & Bandwidth & Resolution & Sensitivity & Source\\
	 & GHz ($\mu m$) & GHz ($\mu m$) & $''$ & mJy & \\
	\midrule
		WC & 1.3675 & 0.144 & $6.1\times7.9$ & $\geq0.14$ & This paper\\
		RACS & 0.8875 & 0.288 & $25\times25^{a}$ & $\geq1$ & \citet{Hale2021}\\
		NVSS & 1.4 & 0.042 & $45\times45$ & $\geq2.2$ & \citet{Condon1998}\\
		{\it WISE W1} & 3.4* & 1.1* & $6.1\times6.1$ & $\geq0.06$ & \citet{Wright2010}\\
		{\it WISE W2} & 4.6* & 1.4* & $6.4\times6.4$ & $\geq0.12$ & \citet{Wright2010}\\
		{\it WISE W3} & 12* & 9.8* & $6.5\times6.5$ & $\geq0.4$ & \citet{Wright2010}\\
	\bottomrule
\end{tabular}
\begin{tablenotes}[para]
    \item[*]Note: Entries are given in terms of wavelength rather than frequency.
    \item[$^{a}$]The initial resolution of RACS observations in this field (CASDA SBID: 12468) are $14''\times14.6''$ which are used to perform any resolved spectral index studies.
\end{tablenotes}
\end{threeparttable}
\end{table*}

\subsection{Radio Data}

The Eridanus field has been included in multiple all-sky radio surveys including NVSS and RACS. The details of the ancillary data used from these observations are presented in Table~\ref{tab:surveytab}.

\subsection{\it WISE}

The Eridanus field was also covered by observations from the {\it WISE} survey. The data is publicly available and obtained from the NASA/IPAC Infrared Science Archive\footnote{https://irsa.ipac.caltech.edu/applications/WISE/} (IRSA). We chose to use {\it WISE} observations due to its high angular resolution, allowing the analysis of resolved properties within galaxies, as well as its coverage and data availability for the entire Eridanus field.

\subsubsection{Flux Measurement}

Integrated {\it WISE} $W1$, $W2$, and $W3$ band flux densities are measured for our Eridanus supergroup galaxy sample using {\tt ProFound} and shown in Table.~\ref{WISEmeasured}. To ensure we are not missing flux from the longer wavelength {\it WISE} bands, we determine the segments used to measure the integrated W2 and W3 flux from the W1 segment map which tend to cover a larger area. IC 1952 contains a background source behind the galactic disk and we are unable to remove its contribution to the total {\it WISE} flux density however the background source accounts for $<5\%$ of the total flux density in each {\it WISE} band as measured by {\tt ProFound}. Errors in the flux density are measured by {\tt ProFound} and added in quadrature to the flux calibration errors of $2.4\%$, $2.8\%$ and $4.5\%$ in W1, W2 and W3 respectively.

\begin{table}[hbt!]
\centering
\begin{threeparttable}
\caption{Measured {\it WISE} fluxes using {\tt ProFound}.}\label{WISEmeasured}
\begin{tabular}{c|c|c|c}
    \toprule
	Common ID & $S_{\rm int}^{\rm W1}$ & $S_{\rm int}^{\rm W2}$ & $S_{\rm int}^{\rm W3}$ \\
	 & (mJy) & (mJy) & (mJy) \\
	\midrule
	IC 1952$^{a}$ & $47.3\pm1.4$ & $28.5\pm0.9$ & $85\pm 5$ \\
	ESO 548-G036 & $28.3\pm0.9$ & $18.8\pm0.7$ & $109\pm 6$ \\
	IC 1953 & $76.5\pm2.2$ & $46.3\pm2.0$ & $218\pm15$ \\
	NGC 1359 & $26.9\pm1.2$ & $15.8\pm0.7$ & $56\pm 4$ \\
	NGC 1367 & $279.0\pm13.0$ & $159.0\pm8.8$ & $88\pm18$ \\
	NGC 1385 & $190.1\pm8.0$ & $130.0\pm4.6$ & $845\pm50$ \\
	NGC 1415 & $179.6\pm4.7$ & $109.8\pm3.7$ & $282\pm15$ \\
	NGC 1422 & $22.6\pm0.7$ & $13.1\pm0.4$ & $29\pm 2$ \\
    \bottomrule
\end{tabular}
\begin{tablenotes}[para]
	\item[]Note. Column (1): Galaxy common ID. Column (2): {\it WISE W1} integrated flux density. Column (3): {\it WISE W2} integrated flux density. Column (4): {\it WISE W3} integrated flux density. 
    \item[$^{a}$]The contribution from the background source is not removed but accounts for $<5\%$ of the total flux density in each {\it WISE} band.
\end{tablenotes}
\end{threeparttable}
\end{table}

We also calculate the Vega magnitude in each {\it WISE} band using:
\begin{equation}
    W_{\rm x} = -2.5\log(\frac{S_{\rm int}^{\rm Wx}}{S_{0}})\bigg.
	\label{eq:mag}
\end{equation}
where $S_{\rm int}$W is the {\it WISE} W1, W2 or W3 integrated flux density in Jy and $S_{0}$ is the {\it WISE} Vega zero magnitude flux density \citet{Wright2010, Jarrett2011} with values of 309.540, 171.787 and 31.674\,Jy for bands $W1$, $W2$ and $W3$ respectively.

\subsubsection{{\it WISE} W3PAH-SFR}

The $W3$ band is an excellent tracer of ISM emission with contributions from PAH emission, nebular emission, and dust heating; however, it also has a contribution from evolved stellar populations. This component of the $W3$ band emission is estimated and removed using the method described in \citet{Cluver2017} whereby the W1 band emission is used as a proxy to determine the stellar emission of which 15.8 percent is determined to be in the W3 band, following \citet{Silva1998}, \citet{Helou2004} and \citet{Jarrett2011}. We subtract 15.8 percent of the W1 band flux from the W3 band flux for both the global and pixel mapped flux densities to determine the contribution from primarily the ISM. We denote these modified W3 flux densities and any other measurements derived from them by using the W3PAH subscript.

The W3PAH luminosity is calculated using:
\begin{equation}
    \bigg(\frac{\rm L_{W3PAH}}{\rm W Hz^{-1}}\bigg)= (9.52\times10^{15})4\pi\,\bigg(\frac{\rm R}{\rm Mpc}\bigg)^{2}\bigg(\frac{\rm S_{\rm int}^{\rm W3PAH}}{\rm mJy}\bigg)
	\label{eq:W3PAHlum}
\end{equation}
with $R$ being the distance to the Eridanus supergroup, $S_{\rm int}^{\rm W3PAH}$ being the integrated W3PAH flux.

The W3PAH-SFR is estimated using the monochromatic W3PAH luminosity ($\nu L_{\rm W3PAH}$) and equation 5 from \citep[][]{Cluver2014}:
\begin{equation}
    \log\bigg(\frac{\rm SFR_{W3PAH}}{\rm M_{\odot}\,yr^{-1}}\bigg)= 1.13\log\bigg(\frac{\nu L_{\rm W3PAH}}{\rm L_{\odot}}\bigg) - 10.24
	\label{eq:Cluver}
\end{equation}
where $\nu$ is the {\it WISE} W3 band central frequency (2.498 $\times 10^{13}$\,Hz) and $L_{\odot}$ = 3.828 $\times 10^{26}$\,W. This relationship has been calibrated against SFR derived from H$\alpha$ observations, which has been shown to be a reliable SFR tracer. Lastly, estimates of the SFR as determined using far-UV (or near-UV, if unavailable) plus {\it WISE} Band 4 24 $\mu$m IR emission from \citet{For2021} are utilised; see \citet{Wang2017} for a detailed description of how these SFRs are calculated.

The use of W3PAH as a SFR tracer (to replace the TIR or FIR measurements traditionally used to measure the IRRC) is primarily due to the availability of high resolution {\it WISE} observations which are used for the resolved IRRC analysis in S~\ref{sec:resolvedirrc} and lack of FIR/TIR data in this region. The {\it WISE} W3 band encompasses a wide range of features within its broad $7-16.5 \mu$m bandwidth most notably the $11.3 \mu$m PAH emission feature as well as nebular lines and continuum emission from warm large grains which in general follows the behaviour of the TIR emission \citep[][]{Li2002}. This emission is thought to be mainly powered by the diffuse interstellar radiation fields produced in star-forming regions \citep[][]{Popescu2011} with PAH fractions being high in regions of active SF suggesting they are produced in molecular clouds \citep[][]{Sandstrom2010}. However PAH's are known to be destroyed by the hard interstellar radiation fields produced by AGNs and massive stars \citep[][]{Smith2007} which is somewhat counteracted by the increase in continuum emission as a result of warm, large grains. Shocks caused by supernovae as well as galactic interactions and mergers are known to destroy PAH's via either thermal or inertial sputtering \citep{draine07, micelotta10, murata17}. \citet{Jarrett2013}, \citet{Cluver2014} and \citet{Cluver2017} find strong correlation between SFRs derived using multiple different well established SFR tracers ($H\alpha$, mid-IR+UV and TIR) and the W3PAH luminosity suggesting its viability as a SFR tracer. \citet{Cluver2014} notes however that the W3PAH-SFR relationship is not expected to hold for low-metallicity environments due to the lack of PAH emission which accounts for $\sim34\%$ of the W3 band emission on average (in their sample). The relationship between W3PAH and TIR luminosities is also not expected to not hold for subregions of galaxies where dust composition, PAH features and dust temperature vary greatly \citep{Cluver2017}. Our sample of galaxies are not significantly starbursting nor do they contain AGN (aside from NGC 1367) and lie within the W3PAH luminosity range where the relationship holds in \citep{Cluver2017}. While we do not have measurements of the metallicity within most of this sample, NGC 1385 was previously found to have a global  $(12+\log(\rm O/H)) > 8.3$ \citep{Williams2021} consistent with abundances used in the \citet{Cluver2017} calibrations. Overall this suggests that use of W3PAH emission as a SFR tracer (and IRRC proxy) in lieu of TIR or FIR emission does not significantly change any conclusions drawn from our global and resolved results.

\subsubsection{Infrared-Radio Correlation}

The IRRC is a relationship that has been observed to be very consistent over wide a range of magnitudes in luminosity and redshift for SFGs \citep{Yun2001, Bell2003, Delhaize2017, Molnar2021} due to its suspected SF-based origin. The IRRC is typically measured using the $q$-parameter:
\begin{equation}
    q_{\lambda} = \log\bigg(\frac{L_{\lambda}}{\rm 3.75\times10^{12}~W}\bigg) - \log\bigg(\frac{L_{1.4}}{\rm W\,Hz^{-1}}\bigg).
	\label{eq:qratio}
\end{equation}
where $L_{\lambda}$ is the IR luminosity at a wavelength ${lambda}$ and $L_{1.4}$ is the total 1.4\,GHz luminosity. We measure $q_{12}$ and $q_{\rm TIR}$ using the W3PAH and TIR luminosities respectively. 

While most analyses of the IRRC use the total or far IR emission to measure $q$, information can still be gleaned from using W3PAH emission as it also traces dust and the ISM, which are heated due to SF and is supported by the tight relationship between W3PAH and TIR emission found in \citet{Cluver2017}:
\begin{equation}
    \log\bigg(\frac{\rm L_{\rm TIR}}{\rm L_{\odot}}\bigg)= (0.889\pm0.018)\log\bigg(\frac{\nu L_{\rm W3PAH}}{\rm L_{\odot}}\bigg) + (2.21\pm0.15).
	\label{eq:LTIR}
\end{equation}
where $\nu L_{\rm W3PAH}$ is the monochromatic W3PAH luminosity.

To be able to observe how the continuum properties of galaxies vary across their disks and whether the group environment in Eridanus affects their physical properties, we measure the resolved IRRC for these eight galaxies. To create the maps which measure the resolved $q_{\rm 12}$ parameter, we first matched the angular resolution of the WALLABY and {\it WISE} W1 and W3 images by convolving the images to a common beam size of 8$''$. This beam corresponds to a distance of 0.8\,kpc and is chosen to maximise the angular resolution and allow observations of the small scale differences within the galactic disks. After removing the sky background component of both {\it WISE} (W1 and W3) images, the W3 images then had their old stellar contribution removed following \cite[][]{Cluver2017} by subtracting $15.8\%$ of the W1 emission on a per pixel basis resulting in what we denote the {\it WISE} W3PAH image. The {\it WISE} W3PAH image was then regridded to a pixel scale of 2$''$ per pixel to match the WALLABY observations. We determine the background and RMS of each WALLABY and W3PAH image using {\tt ProFound} and perform a background subtraction for each W3PAH image. The $q_{12}$ maps are created using Eq.~\ref{eq:qratio} on a per pixel basis where the galaxies show $\geq 3\sigma$ detections in both WALLABY and W3PAH images. These $q_{\rm 12}$ maps are then overlaid on WALLABY HI contours provided by \citet{For2021} to determine whether any features observed relate to properties in the HI observations. We present these images in Fig.~\ref{fig:Murphy1} for each of the eight galaxies in our sample.

\subsection{\emph{g-Band} Optical}

We also make use of co-added \emph{g-band} optical images from the DESI Legacy Imaging Surveys DR8 \citep{Dey2019}, which are based on the Dark Energy Camera Legacy Survey (DECaLS) southern observations, for our sample of Eridanus group galaxies in order to show the extent of radio and IR emission compared to the stellar disks.

\section{Results}
\subsection{Radio Continuum Catalogue}
\label{sec:radiocat}

\begin{table*}[hbt!]
\centering
\begin{threeparttable}
\caption{First five rows of the simplified WC catalogue.}\label{tab:catalogue}
\begin{tabular}{c|c|c|c|c|c|c|c|c|c|c|c}
    \toprule
	 Designation & RAcen & Deccen & S\_peak & S\_int & S\_int & S\_int & S\_int & sky\_RMS & SNR & N100 & R100 \\
   & & & & & \_toterr & \_err & \_err\_cal & \_mean & & & \\
	 & ($^{\circ}$) & ($^{\circ}$) & (mJy/beam) & (mJy) & (mJy) & (mJy) & (mJy) & (mJy) &  & (pixels) & (arcsec) \\
    \midrule
	WC J031926-222623 & 49.859961 & -22.439857 & 8.152 & 11.705 & 1.668 & 0.142 & 1.527 & 0.018 & 32.695 & 59 & 4.683 \\
	WC J031928-222815 & 49.868565 & -22.470727 & 1.628 & 1.516 & 0.321 & 0.123 & 0.198 & 0.018 & 6.761 & 48 & 4.576 \\
	WC J031930-223602 & 49.876814 & -22.600451 & 1.667 & 1.837 & 0.361 & 0.121 & 0.240 & 0.016 & 7.547 & 55 & 5.365 \\
	WC J031931-223941 & 49.877981 & -22.661258 & 1.284 & 1.642 & 0.331 & 0.116 & 0.214 & 0.017 & 5.584 & 47 & 4.949 \\
	WC J031939-223000 & 49.913872 & -22.500039 & 1.043 & 1.140 & 0.239 & 0.090 & 0.149 & 0.015 & 5.114 & 36 & 4.280 \\
    \bottomrule
\end{tabular}
\begin{tablenotes}[para]
    \item[]Note: The columns are described in \ref{sec:catdesc}.
\end{tablenotes}
\end{threeparttable}
\end{table*}

\begin{figure}[hbt!]
	\includegraphics[width=\columnwidth]{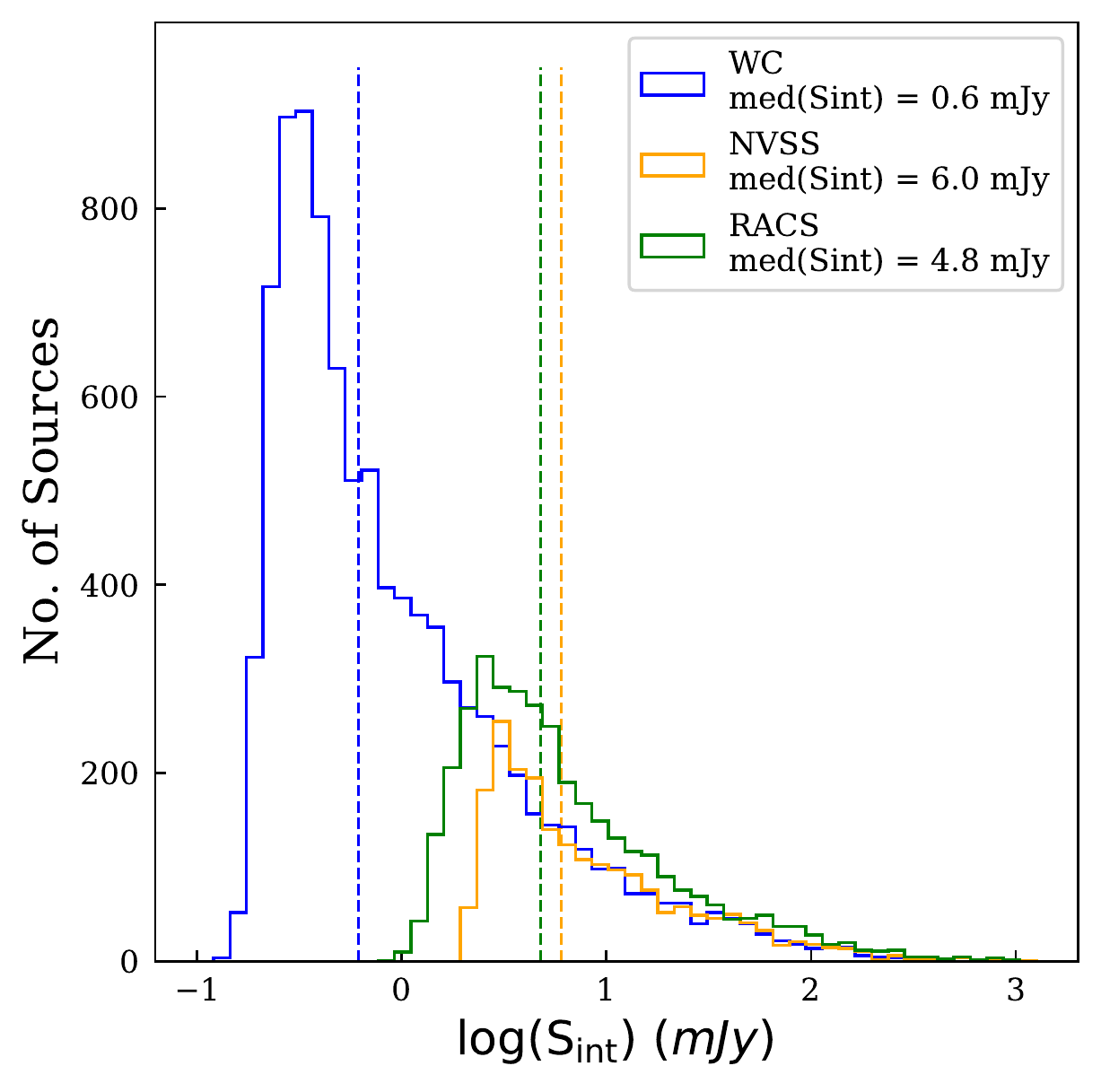}
    \centering
    \caption{Component counts for WALLABY, RACS, and NVSS in the Eridanus field. Flux bins are of width 0.1 dex and vertical dotted lines show the median integrated flux density values of 0.6, 4.8 and 6.0 mJy for WALLABY, RACS and NVSS observations respectively.}
    \label{fig:sourcecount}
\end{figure}

The first five rows of the simplified WC catalogue are shown in Table~\ref{tab:catalogue} with the full catalogue, also containing additional statistics measured by {\tt ProFound}, provided in the online supplementary material. A description for each column included in the extended WC catalogue is provided in~\ref{sec:catdesc} and described in detail in \citet{Robotham2018}. Fig.~\ref{fig:sourcecount} shows the component counts of each survey in the Eridanus field.

\subsubsection{Astrometry}

We assess the astrometry of WC compared to both RACS and NVSS using matched unresolved sources that satisfy the selection criteria in section~\ref{sec:unres}. The RA offset is defined as $\Delta_{\rm RA} = RA_{\rm WC}-RA_{\rm comp}$ where comp refers to the comparison survey and the Dec offset is defined in the same way. We find small mean systematic offsets between WC and NVSS ($\Delta_{\rm RA} = -0.7\pm3.5''$ and $\Delta_{\rm Dec} = -0.1\pm3.5''$), which are constrained to within a couple pixels in the WC observations (2$''$ per pixel) and less than the estimated position error of NVSS sources. Comparisons to RACS also find small mean systematic offsets with $\Delta_{\rm RA} = -0.1\pm1.6''$ and $\Delta_{\rm Dec} = -0.6\pm1.6''$ which are constrained to within one pixel in RACS (2.5$''$ per pixel).

\subsubsection{Flux Offsets}
\label{sec:fo}

Flux density scale comparisons are made between WC and NVSS due to their similar frequencies, which will minimise the impact of errors from the uncertainty in the spectral index. We measure the ratio of integrated flux densities:
\begin{equation}
    R^{\rm a}_{\rm b} = (S_{\rm a}/S_{\rm b})
	\label{eq:R}
\end{equation}
where $S_{a}$ and $S_{b}$ are the integrated flux densities from survey a and b respectively.

We compare the flux densities determined in the WC catalogue to those in the NVSS and RACS catalogues at 1.4\,GHz and 0.888\,GHz respectively. This is to ensure we have a consistent flux scale and to investigate whether the measured spectral index is in agreement with our previous studies of the radio source population. Each of these surveys has a different angular resolution and sensitivity. As such, to ensure that these differences do not impact our comparisons, we restrict comparisons to unresolved, isolated, and high SNR sources as defined in section \ref{sec:unres}.

\begin{figure}[hbt!]
	\includegraphics[width=\columnwidth]{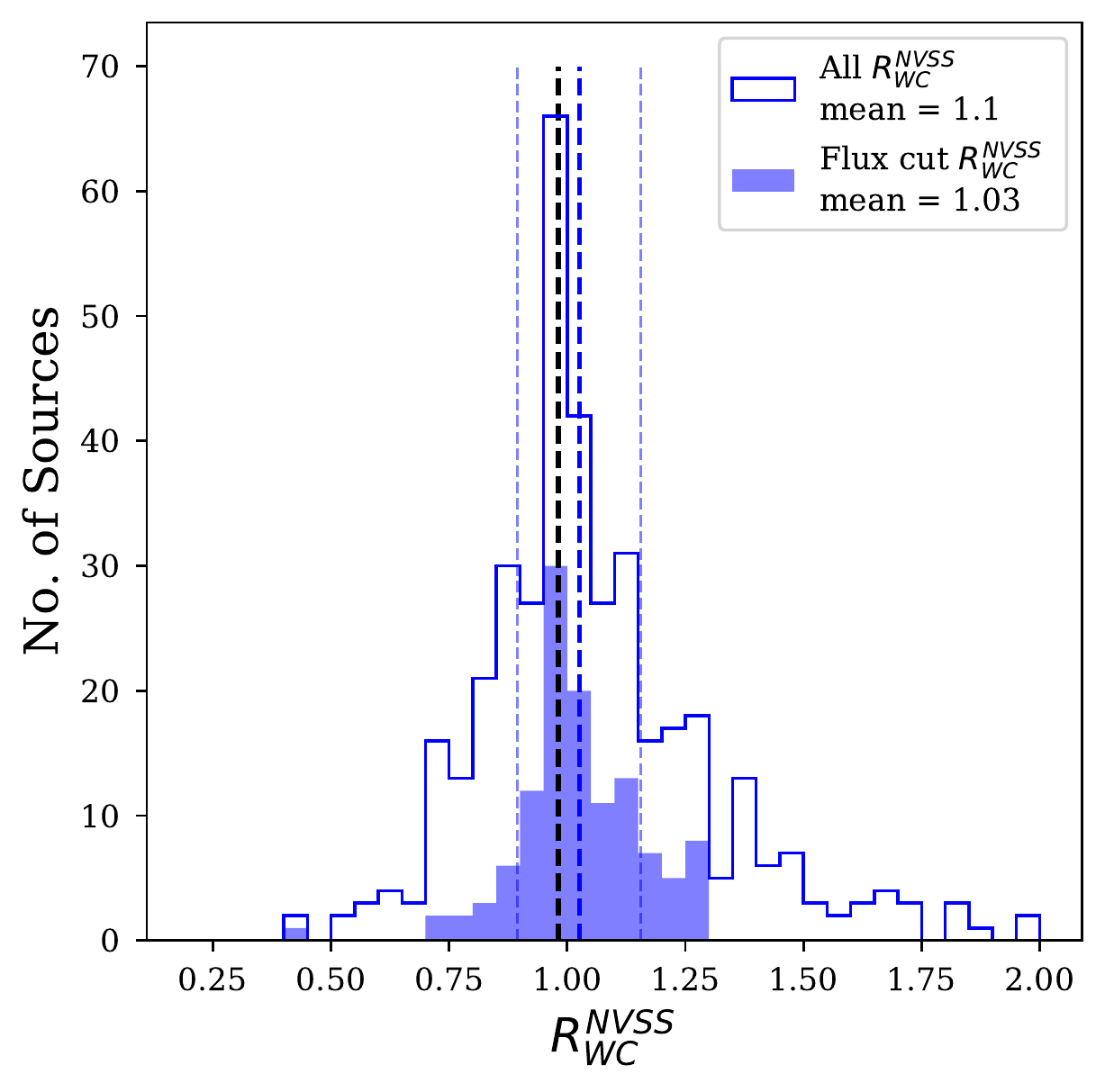}
    \centering
    \caption{Comparison of the integrated flux density ratios between WC and NVSS for the 384 sources matched using the criteria from Section~\ref{sec:unres}. This is shown with (filled) and without (empty) a flux limit of $S_{\rm int}^{\rm NVSS} \geq 6$\,mJy applied and assuming allowed spectral index values of $-2\leq\alpha\leq0.4$. The blue dashed lines indicate the mean flux ratio for the flux limited population and one standard deviation either side whilst the black dashed line indicates the expected flux ratio of 0.98 assuming $\alpha = -0.8$.}
    \label{fig:fluxratios}
\end{figure}

After applying a flux cut of 6\,mJy to NVSS and assuming allowed spectral index values of $-2 \leq \alpha \leq 0.4$ to remove significant outliers and encompass the majority of $\alpha$ values observed \citet{Smolcic2017}, we find the mean flux ratio $R^{\rm NVSS}_{\rm WC} = 1.03 \pm 0.13$ for 122 sources. We therefore conclude that the flux scale for our observation is accurate within the estimated scatter. A $13\%$ error is added in quadrature to the measured WC flux density in order to maintain the flux scale agreement between WC and NVSS throughout this research. However we note that the expected flux ratio between NVSS and WC is 0.98 assuming $\alpha = -0.8$ and so WC likely has a systematic flux deficit of approximately $5\%$, which is also observed in \citet{For2021}. Our source catalogue as well as the RACS catalogue have shown position-dependent flux density variations due to inaccurate primary beam corrections. These issues will be addressed in future observations where instrumental flux calibration will utilise holography beam measurements rather than Gaussian beams. The results are summarised in Fig.~\ref{fig:fluxratios}.

\subsubsection{Spectral Index Analysis}

The spectral index $\alpha$ is used to define the broadband radio emission as a power law of the form $S_{\nu} \sim \nu^{\alpha}$, where $S_{\nu}$ is the flux density at a frequency, $\nu$. Hence, we define $\alpha$ as:
\begin{equation}
    \alpha^{\rm a}_{\rm b} = \frac{\log(S_{\rm a}/S_{\rm b})}{
    \log(\nu_{\rm a}/\nu_{\rm b})}.
	\label{eq:alpha}
\end{equation}
where $\nu_{a}$ and $\nu_{b}$ are the frequencies of survey a and b respectively. $\alpha$ is typically found to have values between $-0.7$ and $-0.8$ for SFGs in the synchrotron dominated regime ($\nu \leq 3 \,GHz$) \citep{Condon1992, Smolcic2017}. We also make comparisons after performing flux cuts that are chosen to prevent any bias in $\alpha$, which would bias the spectral indices to higher or lower values based on sensitivity limits and frequencies of the comparison surveys.

We measure the spectral indices between WC and RACS as well as NVSS and RACS (for sources common to WC) with results summarised in Table~\ref{fluxcomptable}. The spectral index values between WC and RACS imply a steeper than normal spectral index; however, given the significant scatter in measured values, the expected value of -0.8 is recovered. Interestingly, $\alpha^{\rm NVSS}_{\rm RACS}$ is measured to be flatter than $\alpha^{\rm WC}_{\rm RACS}$ without taking into account the large error budget. The steeper than normal spectral index between WC and RACS can be explained by the compounding systematic flux offsets where WC has an approximately $5\%$ flux deficit and RACS has a slight excess, as suggested by the spectral index analysis of RACS \citep[see Section 5.2.5 of][]{Hale2021}. The agreement between RACS and NVSS hints that the primary cause for discrepancies in the comparisons is the slight WC flux deficit, measured to be $\alpha^{\rm WC}_{\rm RACS} = -0.91 \pm 0.41$  as compared to $\alpha^{\rm NVSS}_{\rm RACS} = -0.80 \pm 0.39$ for the same 122 sources. The spectral index values between WC and RACS are measured to be steeper when looking at the entire sample compared to after applying a flux cut suggesting that either WC is measuring less flux or RACS is measuring excess flux for faint sources. We believe this is likely due to imaging differences between WC and RACS including different weightings, and RACS convolution to a 25$''$ FWHM beam. We note that these discrepancies are not statistically significant due to the magnitude of uncertainty in spectral indices between WC, RACS, and NVSS. These findings suggest that future comparisons to the WC and RACS catalogues will exhibit systematic flux discrepancies; however, future improvements to ASKAP processing will likely minimise such discrepancies.

\begin{table*}[hbt!]
\centering
\begin{threeparttable}
\caption{Flux offsets and spectral indices between catalogues.}\label{fluxcomptable}
	\begin{tabular}{c|c|c|c|c}
    \toprule
	Matched Catalogues & Sample & No. of Sources  & Median $\alpha^{a}_{b}$ & Mean $\alpha^{a}_{b}$ \\
	 &  &  & 16th and 84th Percentiles & $\pm~\mbox{Std(SE)}$ \\
	\midrule
	RACS-WC & All & 1418 & $-1.24^{+0.71}_{-0.65}$ & $-1.26\pm~0.77$ \\
	 & 3\,mJy RACS cut & 697 & $-1.11^{+0.46}_{-0.48}$ & $-1.09\pm~0.47$ \\
	RACS-NVSS (WC) & All & 384 & $-0.92^{+0.64}_{-0.45}$ & $-0.82\pm~0.67$ \\
	 & 6\,mJy NVSS cut & 122 & $-0.87^{+0.45}_{-0.30}$ & $-0.80\pm~0.39$ \\
	RACS-WC (NVSS) & All & 384 & $-1.00^{+0.42}_{-0.50}$ & $-0.94\pm~0.55$ \\
	 & 6\,mJy NVSS cut & 122 & $-0.96^{+0.31}_{-0.39}$ & $-0.91\pm~0.41$ \\
	RACS-WC & Eridanus & $5^{c}$ &  & $-1.34\pm~(0.12)$ \\
	RACS-NVSS & Eridanus & $6^{c}$ & & $-0.87\pm~(0.11)$ \\
   \bottomrule
\end{tabular}
 \begin{tablenotes}[para]
	\item[]Note: Column (1): Catalogues compared in a-b format. Column (2): Sample used. All flux cuts also include cuts between $-2 \leq \alpha^{a}_{b} \leq 0.4$. Column (3): Number of sources in the sample from catalogue $a$ (top) and catalogue $b$ (bottom). Column (4): Median spectral index for sample with 16th and 84th percentiles. Column (5): Mean spectral index for sample with the standard deviation (standard error) about the mean. 
    \item[$^{c}$]Outliers have been removed.
\end{tablenotes}
\end{threeparttable}
\end{table*}

\subsection{Radio Continuum Properties of Eridanus Supergroup}
Table~\ref{tab:WCmeasured} showed the measured radio continuum fluxes for Eridanus supergroup members. Overall we see agreement between the WC, NVSS and \citet{Omar2005} $\sim1.4\,GHz$ integrated flux densities for ESO 548-G036 and NGC 1415, which exhibit compact radio emission. The higher resolution afforded by WC observations allows us to discern that, for IC 1952, a majority of the radio emission is due to a background radio source behind the disk of IC 1952. After subtracting this background contribution, the measured flux density of IC 1952 is $\sim$6 times lower than previous surveys that were unable to separate and remove this background contribution. WC measures a flux deficit for IC 1953 and NGC 1385, which both have extended diffuse emission that is likely resolved out. We also measure a flux deficit for NGC 1359 due to loss of sensitivity at the edge of the field. Our flux for NGC 1367 agrees with NVSS; however, \citet{Omar2005} measures $\sim50\%$ more. WC has higher sensitivity to point sources than previous surveys allowing us to measure the faint radio continuum emission from the nucleus of NGC 1422 which was previously undetected. Column 1 of Fig.~\ref{fig:Murphy1}, shows the WALLABY continuum emission contours overlaid on SDSS \emph{g-}band optical images for each of these eight galaxies.

\subsubsection{Global Spectral Indices}

The spectral index values for each individual galaxy can be seen in Table~\ref{tab:WCDerived}. Measuring the global spectral index values for these galaxies between WC and RACS as well as RACS and NVSS, we find the mean of the derived values of $\alpha^{\rm WC}_{\rm RACS} = -1.34 \pm 0.12$ and $\alpha^{\rm NVSS}_{\rm RACS} = -0.87 \pm 0.11$ respectively, where errors are the standard error of the mean. We remove outliers IC 1953 and IC 1952, the latter being contaminated by a background radio source. IC 1953 has flux deficits in WC and RACS, likely due to resolving out extended radio flux, resulting in $\alpha > 0.9$ in both cases. The spectral index value we find for the Eridanus sample is steeper than the $\sim-0.8$ generally found for SFGs \citep{Condon1992, An2021} whilst $\alpha^{\rm NVSS}_{\rm RACS}$ is in agreement with this value. Considering the steeper spectral index between WC and RACS compared to RACS and NVSS for our nearby galaxy sample  we assume that WC is missing some extended flux. Overall, due to the small population biases and systematic flux offsets between WC and RACS, we are unable to observe any obvious impact that environmental interactions have on the global radio spectral indices.

\begin{table}[hbt!]
\centering
\begin{threeparttable}
\caption{Derived radio continuum properties of Eridanus supergroup members.}\label{tab:WCDerived}
\begin{tabular}{c|c|c|c}
    \toprule
	Common ID & L$_{WC}$ & $\alpha^{WC}_{RACS}$ & $\alpha^{NVSS}_{RACS}$ \\
	 & $\times 10^{20}$ (W Hz$^{-1}$) &  & \\
    \midrule
	IC 1952 & $0.51\pm0.15$ & $-4.79\pm0.80$ & $-1.16\pm0.47$ \\
	ESO 548-G036 & $4.19\pm0.57$ & $-1.20\pm0.43$ & $-1.23\pm0.31$ \\
	IC 1953 & $3.66\pm0.75$ & $0.91\pm0.78$ & $2.24\pm0.69$ \\
	NGC 1359 & $9.98\pm1.89$ & $-1.74\pm0.47$ & $-0.56\pm0.24$ \\
	NGC 1367 & $6.96\pm0.95$ & $-1.51\pm0.39$ & $-1.09\pm0.34$ \\
	NGC 1385 & $75.06\pm9.88$ & $-1.26\pm0.35$ & $-0.70\pm0.18$ \\
	NGC 1415 & $12.36\pm1.64$ & $-0.96\pm0.37$ & $-0.75\pm0.22$ \\
	NGC 1422 & $0.47\pm0.09$ &  &  \\
   \bottomrule
\end{tabular}
\begin{tablenotes}[para]
	\item[]Note. Column (1): Galaxy common ID. Column (2): WC luminosity assuming a distance of 21\,Mpc. Column (3): Spectral index between WC and RACS. Column (4): Spectral index between NVSS and RACS.
\end{tablenotes}
\end{threeparttable}
\end{table}

\subsubsection{Resolved Spectral Index Properties of NGC 1385}

NGC 1385 is the brightest galaxy in our sample and is almost face-on with resolved radio continuum observations also available in RACS at 888\,MHz \citep{Hale2021}. This allows us to investigate the resolved spectral index properties and investigate the environmental impact on CR electron transport. As flux is missing in the WC observations of NGC 1385 we primarily perform this as a qualitative analysis to observe whether a significant gradient is observed across the galactic disk. To generate these maps we convolve the WALLABY data to a matching $14.6 \times 14''$ beam FWHM with RACS observations of the Eridanus field. We then regrid to a common pixel size of $2.5''$ and positionally match the images. The spectral index is calculated using Eq.~\ref{eq:alpha} for each pixel with a SNR above $10\sigma$ in both images. We expect to find a flatter spectral index in SF regions and a steeper spectral index in regions where CR electron transport has occurred.

\begin{figure}[hbt!]
	\includegraphics[width=\columnwidth]{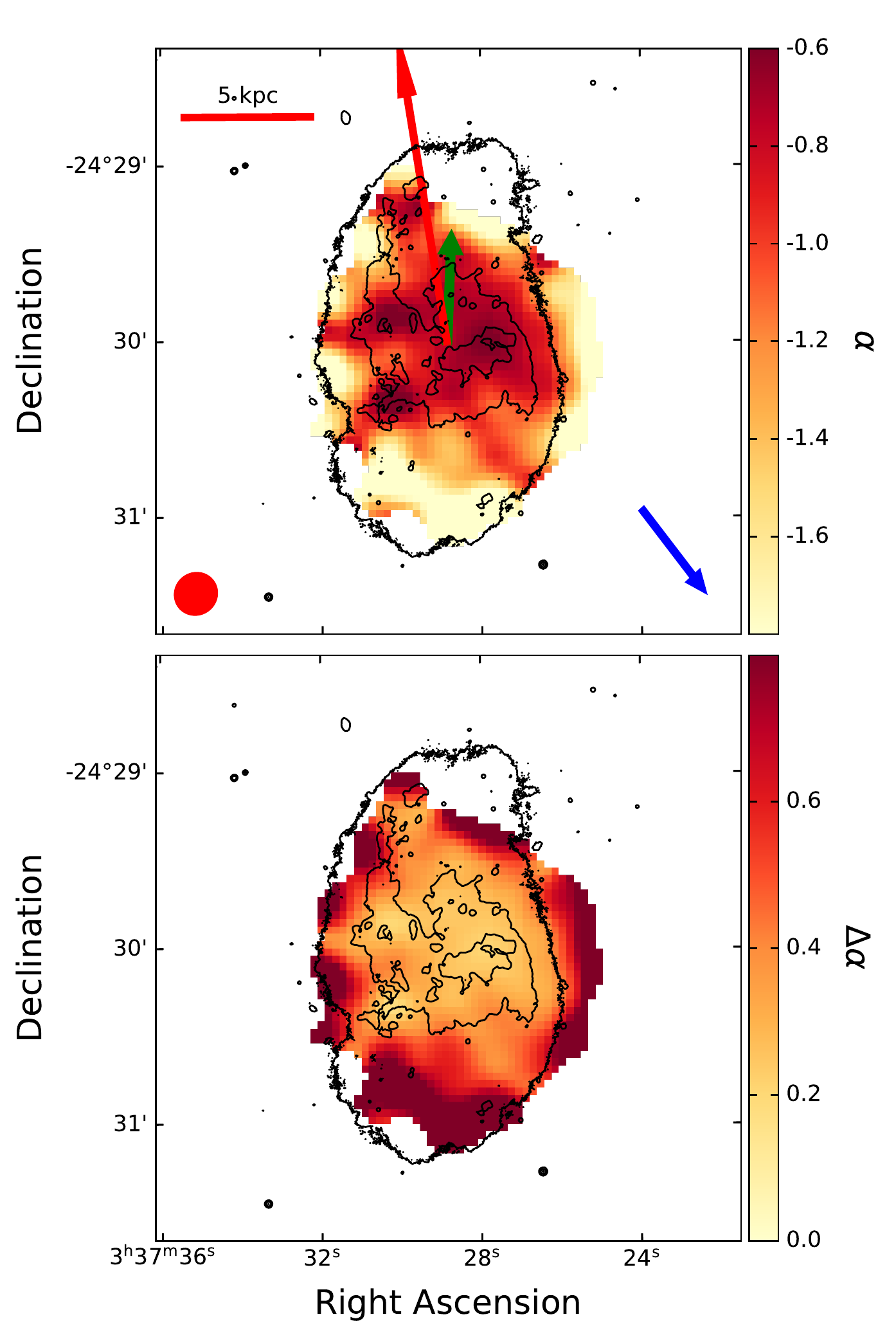}
    \centering
    \caption{Top: Spectral index map of NGC 1385 between WALLABY at 1.37\,GHz and RACS at 0.88\,GHz. Contours from $g$-band SDSS images are overlaid to show the extent and brightest regions of the galaxy. The red bar indicates a scale length of 5\,kpc whilst the red circle indicate the RACS FWHM beam size of $14.6 \times 14''$. The red arrow indicates 1/50th of the distance to the group central galaxy NGC 1395 and the green arrow indicates 1/50th of the distance towards Dark cloud 2 \citep{Wong2021}. The blue arrow indicates the likely direction of CR electron streaming and we note a slightly flatter spectral index to the south-west of the galactic centre (ignoring the outer edges of the galaxy). Bottom: The estimated error in the spectral index map of NGC 1385.}
    \label{fig:SImap1385}
\end{figure}

The spectral index map between WC and RACS for NGC 1385 (Fig.~\ref{fig:SImap1385}) is quite steep with a median $\alpha^{\rm WC}_{\rm RACS} = -1.10$ and mean $\alpha^{\rm WC}_{\rm RACS} = -1.23$, which agree with the global value of $\alpha^{\rm WC}_{\rm RACS} = -1.26 \pm 0.35$. The lower values measured throughout the galaxy can be partly explained by the missing flux in WC and flux excess in RACS compounding their systematics. We see some structure in the spectral index map with a flatter spectral index being measured in the nucleus and star forming regions in the spiral arms where the injection spectral index is measured to be $\alpha^{\rm WC}_{\rm RACS} \sim -0.7$ as compared to $\alpha_{\rm synch} = -0.65\pm0.1$ measured in \citet{Tabatabaei2013} for NGC 6946. There appears to be a slight gradient of steepening spectral index across the galactic disk from north-east to south-west (as denoted by the blue arrow) coinciding with the HI and IRRC features suggesting that evidence of CR electron streaming may be seen in the spectral index map of NGC 1385.

\subsection{Star-Formation Properties of the Eridanus Supergroup}
Radio continuum emission has been shown to be an excellent tracer of recent star-formation due to its dependence on massive short-lived stars and its dust unbiased nature. To investigate whether the measured global radio SFRs for these galaxies are consistent with SFRs measured using other multiwavelength methods we use the relationship from \citet{Molnar2021}:
\begin{equation}
    \log\bigg(\frac{\rm SFR_{1.4}^{\rm Molnar}}{\rm M_{\odot}\,yr^{-1}}\bigg)= (0.823\pm0.009)\log\bigg(\frac{L_{1.4}}{\rm W\,Hz^{-1}}\bigg) - (17.5\pm0.2)
	\label{eq:Molnar}
\end{equation}
where $L_{1.4}$ is the 1.4\,GHz radio luminosity in ${\rm W~Hz^{-1}}$ and is calibrated against the $q_{\rm TIR}$ (see Eq.~\ref{eq:qratio}) derived SFR to estimate the radio SFR. The WC flux density is calibrated against NVSS and as such the 1.4\,GHz luminosity is calculated using:
\begin{equation}
    \bigg(\frac{\rm L_{\rm WC}}{\rm W\,Hz^{-1}}\bigg)=(9.52\times10^{15})4\pi\,\bigg(\frac{\rm R}{\rm Mpc}\bigg)^{2}\bigg(\frac{S_{\rm int}^{\rm WC}}{\rm mJy}\bigg)
	\label{eq:luminosity}
\end{equation}
where R is the distance to the Eridanus supergroup in Mpc and $S_{\rm int}$ is the integrated flux density. We do not apply K-corrections to the luminosity due to the small distance of the Eridanus supergroup.

We compare the radio SF properties of Eridanus supergroup members to those using other multiwavelength methods. In \citet{For2021} for both IC 1953 and ESO 548-G036 only the W4-SFRs are calculated due to UV contamination and are assumed to be lower limits. All radio and UV+W4 derived SFR measurements are estimated using a Chabrier initial mass function (IMF) \citep{Chabrier2003}. The W3PAH-SFR relationship from \citet{Cluver2014} relies on the BG03 IMF \citep{Baldry2003} which has been shown to be accurate for low SFR systems \citep{Cluver2014}. Conversion of the BG03 IMF into a Chabrier IMF causes the SFR relationship in \citet{Cluver2014} to overestimate the SFRs by a factor of $\sim2$ by construction due to their calibration.

\begin{figure*}[hbt!]
	\includegraphics[width=\textwidth]{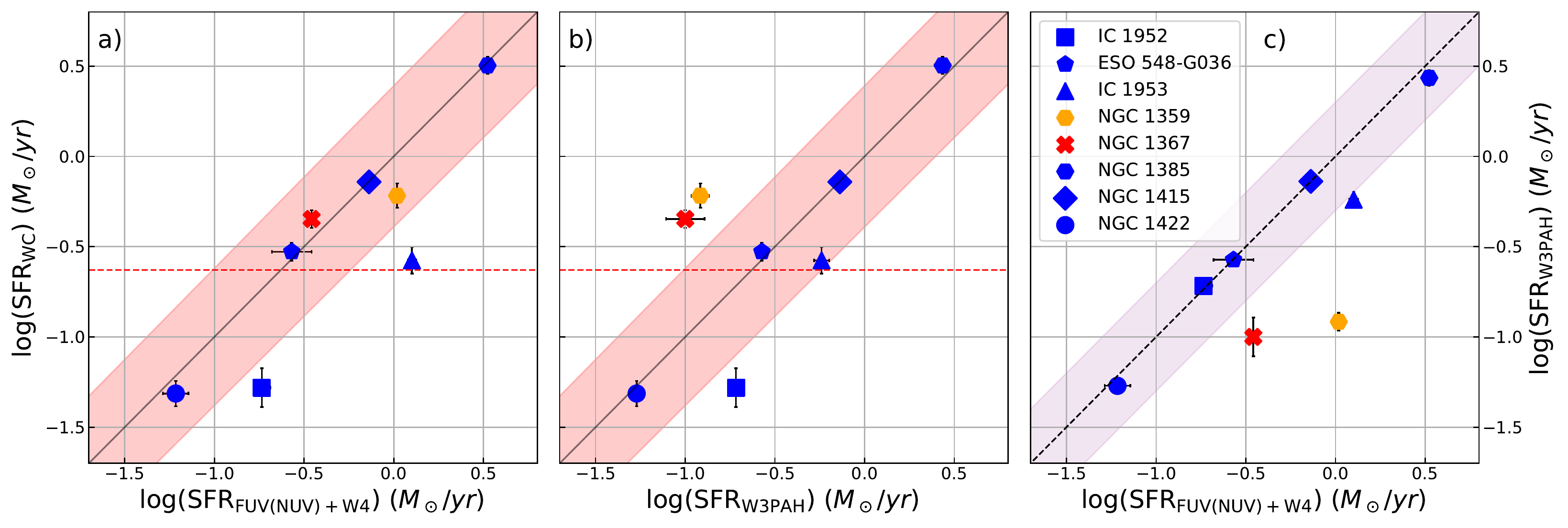}
    \caption[width=\textwidth]{Comparison between global SFRs measured for the eight Eridanus supergroup galaxies. a) Comparison between radio (WC) and FUV(NUV)+$W4$ derived SFRs. b) Comparison between radio (WC) and W3PAH derived SFRs. c) Comparison between W3PAH and FUV(NUV)+$W4$ derived SFRs. $SFR_{\rm WC}$ is the 1.4 GHz SFR as determined using eq.~\ref{eq:Molnar} \citep[][]{Molnar2021}. Error bars result from measured flux uncertainties in {\tt ProFound} and calibration errors. Red shaded region indicates the 1$\sigma$ scatter in the radio-SFR relationship derived in \citet{Molnar2021} with their luminosity limit indicated by the horizontal red line. Blue shaded region indicates 0.2 dex scatter. Both ESO 548-G036 and IC 1953 are assumed to be lower limits for FUV(NUV)+$W4$ derived SFRs due to contamination allowing only W4-SFR to be determined.}
    \label{fig:SFR}
\end{figure*}

We see agreement between the SFR estimates in Fig.~\ref{fig:SFR} and Table \ref{tab:SFRtab}, especially for the sources with higher SFRs, with all but IC 1952 and IC 1953 being within the scatter of the radio-SFR relationship (Eq.~\ref{eq:Molnar}). Fig.~\ref{fig:SFR}b reveals discrepancies between the radio and IR emission for some of the Eridanus galaxies, from which details about their physical properties can be inferred. Most of the galaxies show an agreement within the scatter between their SFRs as determined by their W3PAH and radio flux densities. NGC 1359 appears to have a deficit of W3PAH flux in comparison to the radio flux. NGC 1367 also has significant excess radio continuum emission relative to the W3PAH SFR but is comparable to the UV+W4 SFR. IC 1952 lies below the 1:1 line for both comparisons involving the radio SFR due to the removal of the strong background radio source, which results in a loss of a majority of the measured radio flux from IC 1952. This background source however does not contribute significantly to the IR or UV emission. Lastly, we see excellent agreement between the W3PAH and UV+W4 derived SFRs for all but NGC 1367 and NGC 1359 providing further evidence that we have a deficit of W3PAH emission in these galaxies as well as a radio deficit for IC 1952.

\begin{table*}[hbt!]
\centering
\begin{threeparttable}
\caption{Derived SFR properties of Eridanus supergroup members.}\label{tab:SFRtab}
	\begin{tabular}{c|c|c|c|c|c|c}
    \toprule
	Common ID & $SFR_{\rm WC}^{\rm Molnar}$ & $SFR_{\rm UV+W4}$ &$SFR_{W3PAH}$ & $\log(\Sigma(SFR_{\rm WC}))$ & $\log(\Sigma(gas))$ & Area \\
	 & ($M_{\odot} yr^{-1}$) & ($M_{\odot} yr^{-1}$) & ($M_{\odot} yr^{-1}$) & ($M_{\odot} yr^{-1} kpc^{-2}$) & ($M_{\odot} yr^{-1} pc^{-2}$) & ($kpc^2$) \\
	\midrule
	IC 1952 & $0.05\pm0.01$ & $0.18\pm0.02$ & $0.19\pm0.02$ & $-2.59\pm0.05$ & $0.94\pm0.01$ & 28 \\
	ESO 548-G036 & $0.30\pm0.03$ & $>0.27\pm0.07^{b}$ & $0.27\pm0.02$ & $-2.03\pm0.05$ & $0.38\pm0.01$ & 32.6 \\
	IC 1953 & $0.27\pm0.04$ & $>1.26\pm0.07^{b}$ & $0.58\pm0.06$ & $-1.92\pm0.05^{c}$ & $0.25\pm0.04$ & 8.7 \\
	NGC 1359 & $0.61\pm0.09$ & $1.04\pm0.07$ & $0.12\pm0.01$ & $>-2.60\pm0.05^{d}$ & $0.98\pm0.002$ & 225.5 \\
	NGC 1367 & $0.45\pm0.05^{a}$ & $0.35\pm0.02$ & $0.10\pm0.02$ &  &  &  \\
	NGC 1385 & $3.19\pm0.35$ & $3.33\pm0.28$ & $2.72\pm0.28$ & $-2.04\pm0.05$ & $0.89\pm0.001$ & 358 \\
	NGC 1415 & $0.72\pm0.08$ & $0.73\pm0.07$ & $0.73\pm0.06$ & $-1.56\pm0.05$ & $0.46\pm0.01$ & 28.1 \\
	NGC 1422 & $0.05\pm0.01$ & $0.06\pm0.01$ & $0.05\pm0.005$ & $-2.20\pm0.05$ & $0.78\pm0.07$ & 5.2 \\
    \bottomrule
\end{tabular}
\begin{tablenotes}[para]
	\item[]Note. Column (1): Galaxy common ID. Column (2): Total WC SFR. Column (3): $UV+W4$ SFR determined in \citet{For2021}. Column (4): Total W3PAH SFR. Column (5): WC SFR surface density of the WC selected star-forming regions of each galaxy. Column (6): Gas surface density of the WC selected star-forming regions of each galaxy. Column (7): Area of star-forming ellipse used to calculate surface densities. 
    \item[$^{a}$]WC emission is entirely AGN based however we provide a SFR based on the global emission for comparison. 
    \item[$^{b}$]Considered as lower limits due to UV contamination. 
    \item[$^{c}$]We only consider the nuclear starburst region. 
    \item[$^{d}$]Due to the irregular shape of NGC 1359 WC emission being improperly measured by an ellipse this value is considered as a lower limit.
\end{tablenotes}
\end{threeparttable}
\end{table*}

We also examine where each of these galaxies lie on the Kennicutt-Schmidt law \citep[KS-law;][]{Kennicutt1998}:
\begin{equation}
    \Sigma_{\rm SFR} \propto \Sigma^{N}_{\rm gas}.
	\label{eq:KSeq}
\end{equation}
where $\Sigma_{\rm SFR}$ is the SFR surface density and $\Sigma_{\rm gas}$ is the cold gas mass surface density, which has a power-law index of N = 1.4 based on observations of samples of spiral and starburst galaxies. We measure the integrated flux densities of WC and WALLABY HI in an ellipse chosen to encapsulate the WC emission of each galaxy. This method ensures we accurately estimate the instantaneous SFR surface density. We convert the WC integrated flux density into a SFR using Eq.~\ref{eq:Molnar} and then calculate the SFR surface density for each galaxy. The HI integrated flux density is converted into an HI gas mass using equation 6 from \citet{For2021}, and using a mass correction for helium of 1.35 \citep{Obreschkow2016, Murugeshan2021}, the cold gas mass surface density is calculated for each galaxy.

\begin{figure}[hbt!]
	\includegraphics[width=\columnwidth]{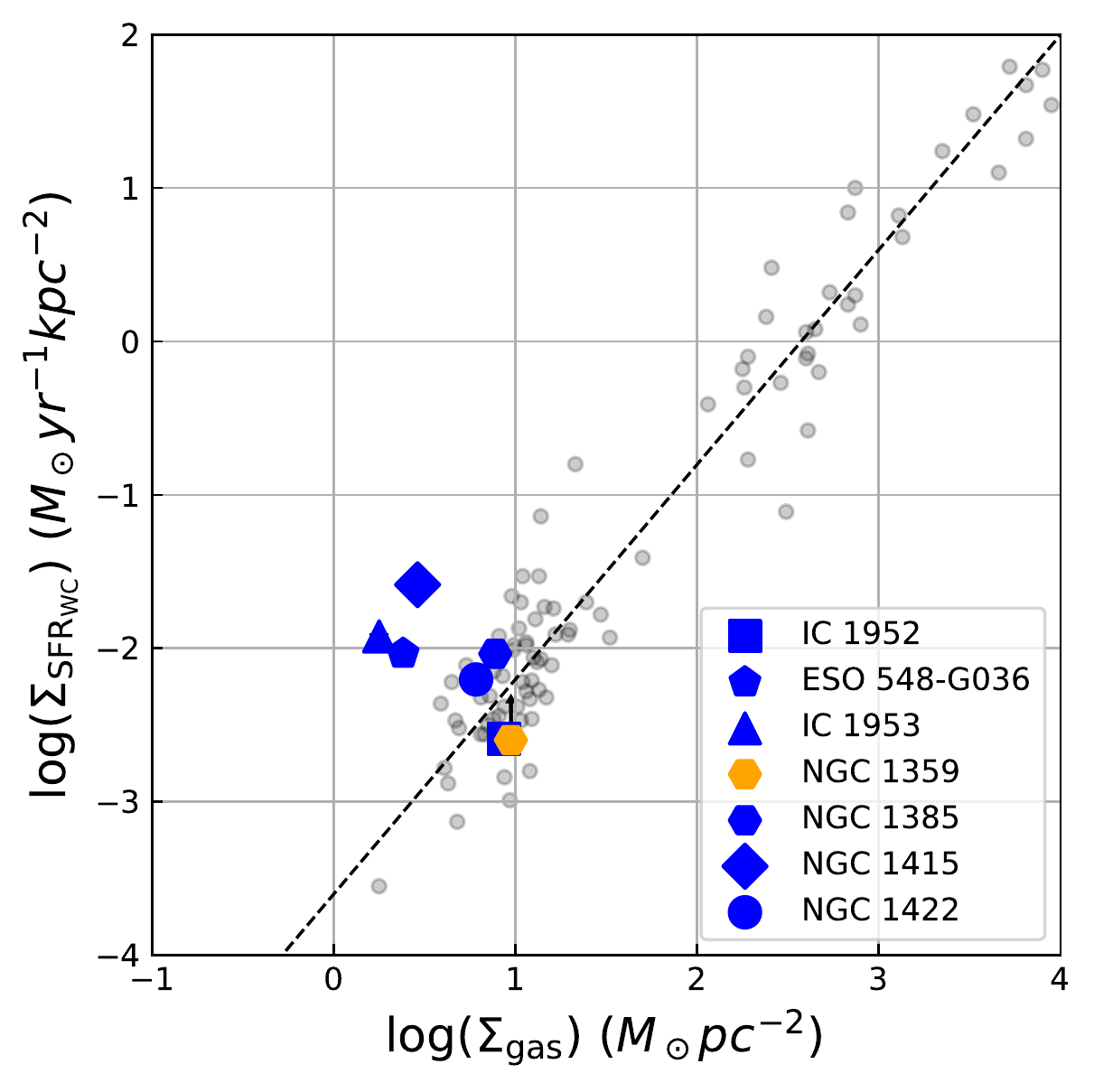}
    \centering
    \caption{Kennicutt-Schmidt law for the sample of Eridanus galaxies. NGC 1359 is assumed to be at the lower limit of its SFR density due to its irregular morphology. Grey points indicate the sample of normal disk and starburst galaxies from \cite{Kennicutt1998} with the dashed line indicating a power-law index of 1.4.}
    \label{fig:KSplot}
\end{figure}

We find that most galaxies lie within the scatter of the KS-law with a few notable outliers (see Fig.\ref{fig:KSplot}). Most galaxies lie above the KS-law suggesting that there is an overall gas deficiency as compared to their radio SFRs for these group member galaxies which agrees with the findings of \citet{For2021} and \citet{Murugeshan2021}. None of these galaxies have $log(\Sigma_{\rm SFR}) \geq -1$ putting them below the typical threshold for starburst galaxies. IC 1953, ESO 548-G036 and NGC 1415 sit well above the KS-law indicating an enhanced SFR density likely due to their compact radio continuum morphology or lack of sensitivity to faint extended radio continuum emission. Both IC 1953 and NGC 1415 are also known to be undergoing nuclear starbursts.

\begin{figure}[hbt!]
	\includegraphics[width=\columnwidth]{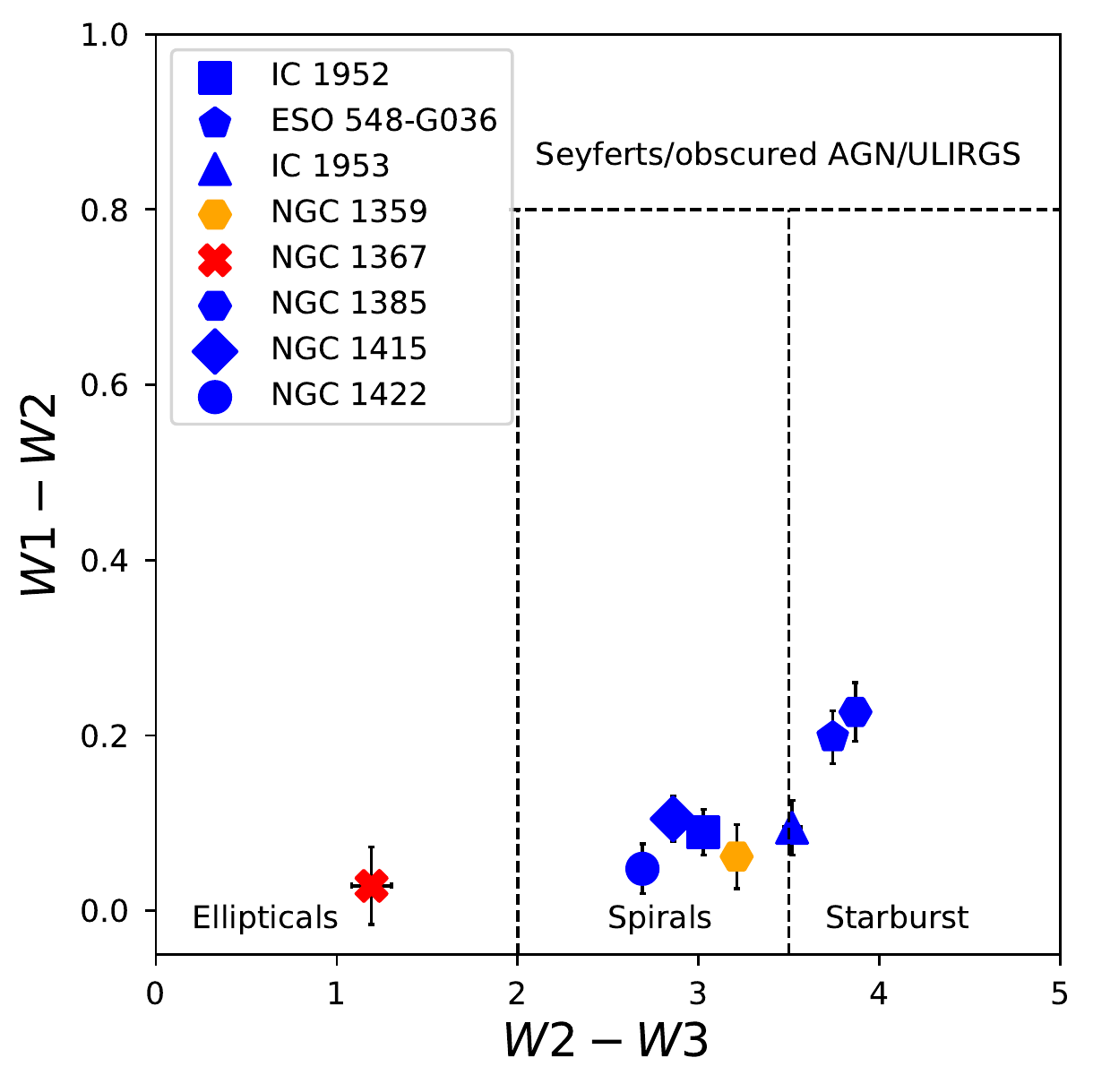}
    \caption{{\it WISE} colour-colour diagram \citet{Jarrett2013} of the sample of eight Eridanus galaxies. Early type galaxies with low SFR are expected to occupy the bottom left of this diagram (W2 - W3 < 2), starbursting disk galaxies lie on the right side of this diagram (W2 - W3 > 3.5) with normal star-forming spirals lying between these regions. At above W1 - W2 = 0.8 mid infrared emission is dominated by heating from dusty AGNs.}
    \label{fig:WISEcmd}
\end{figure}

{\it WISE} flux measurements allow us to see where these galaxies lie on the {\it WISE} colour-colour diagram which can help separate highly efficient accreting AGN and starburst galaxies from passively evolving ones \citep{Jarrett2013, Cluver2014}.Fig.~\ref{fig:WISEcmd} shows that all of the galaxies in this sample lie in the $W1-W2 \leq 0.3$  colour space suggesting that these galaxies are passively evolving and are not dominated by highly AGN ($>1\%$ Eddington). NGC 1367 occupies the lower left of the plot where low SFR early-type galaxies usually lie. Most of the other galaxies lie within the region usually occupied by low SFR spiral galaxies with NGC 1385 and ESO 548-G036 displaying slightly higher $W2-W3$ colour suggesting higher levels of SF activity. Interestingly, despite the observed post-merger starburst activity of NGC 1359, it lies in the same region as the other lower SFR spirals. When compared to the KS-law diagram (Fig. \ref{fig:KSplot}) we see agreement between the classifications for most galaxies which are categorised as star-forming spirals. NGC 1385 and ESO 548-G036 are likely undergoing increased SF activity and NGC 1367's radio emission is a result of AGN processes.

\subsection{Global IRRC Properties of the Eridanus Supergroup}
In terms of the $q_{12}$ values, we see an agreement with the general trends shown in \citet{Molnar2021, An2021} whereby $q$ decreases with increasing radio or TIR luminosity. We also observe that NGC 1367 lies below the other galaxies due to the AGN mechanism being responsible for excess radio emission in the form of radio jets. NGC 1359 lies in the same region as NGC 1367 despite being a SFG as a result of low W3PAH emission. The trend we see here appears to be steeper than the trends generally observed for SFGs in \citet{Molnar2021} eq. 16 \& 18. When comparing the $q_{\rm 12}$ values to $q_{\rm FIR}$ values measured in \citet{Omar2005} we find the values follow the same trend for each galaxy with $q_{\rm FIR}$ being offset to higher values by $\sim0.6$ as expected due to the higher FIR luminosity. The exception to this is IC 1952 in which we remove the background radio source which isn't individually resolved and removed in \citet{Omar2005}.

\begin{figure*}[hbt!]
	\includegraphics[width=\textwidth]{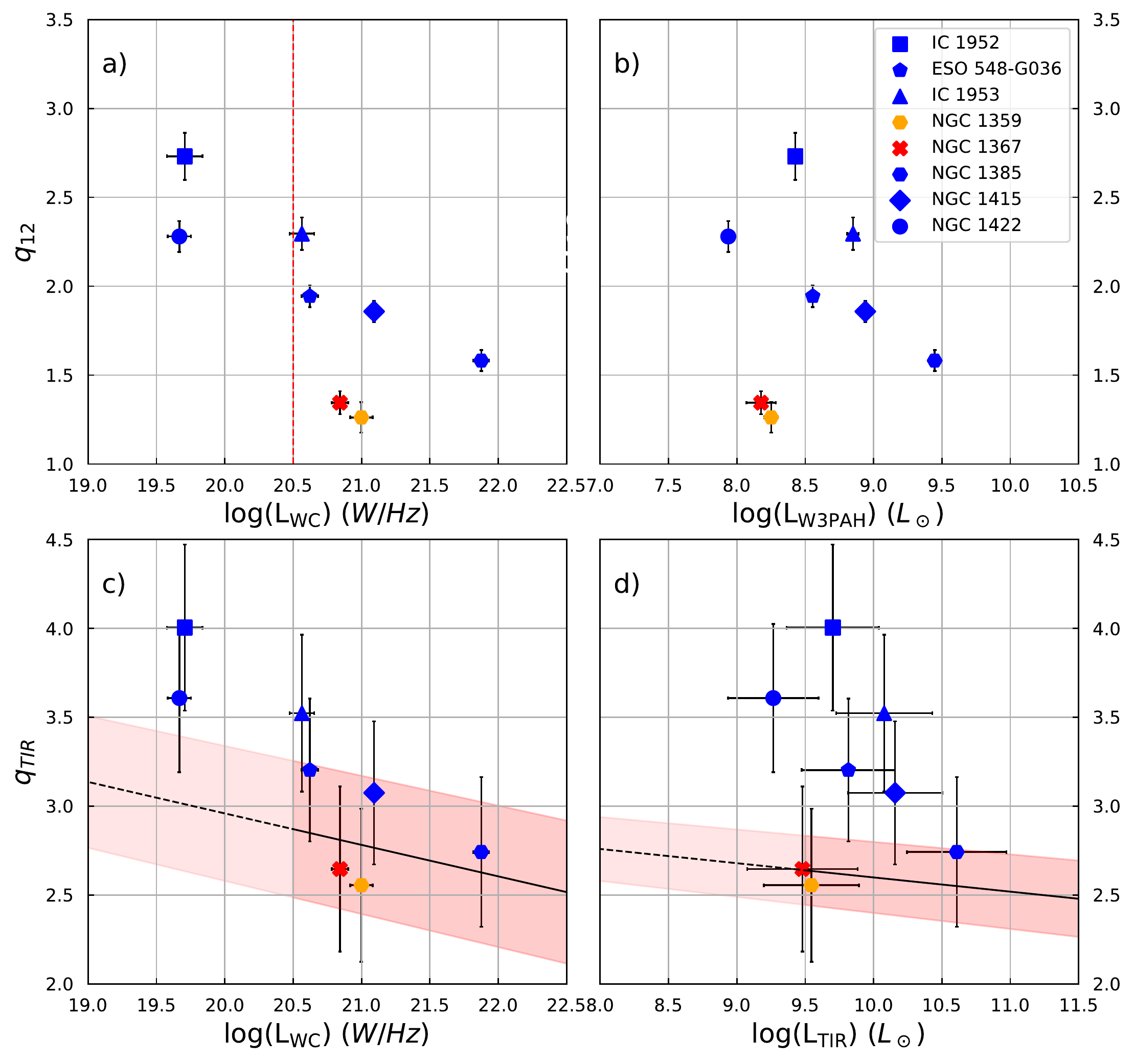}
    \caption[width=\textwidth]{Top: The global $q_{12}$ value measured for each galaxy compared to their a) WC luminosity and b) W3PAH luminosity. The estimated $q_{\rm TIR}$ values determined using the relationship from \citet{Cluver2017} compared to their c) WC luminosity and d) estimated TIR luminosity. The error bars are determined using error propagation of the flux uncertainties as well as the addition in quadrature of the scatter in the estimate for TIR luminosity in \citet{Cluver2017}. The black line with red scatter indicates the relationships for $q_{\rm TIR}$ measured in \citet{Molnar2021} for SFGs. The red vertical line and dashed portion of the \citep[][]{Molnar2021} relationship indicates the luminosity below which data is extrapolated.}
    \label{fig:qplot}
\end{figure*}

Discrepancies can partly be explained by IC 1952 and IC 1953 having a radio flux deficit due to removal of a background AGN source and missing extended flux respectively which would boost their $q_{\rm 12}$ values. We also note that these Eridanus galaxies have low SFRs and generally lie on or slightly below the luminosity limit where most IRRC relationships are measured and so there is likely a larger amount of inherent scatter at these luminosities. After converting the W3PAH emission into an estimated TIR emission, we investigate whether the W3PAH emission is a good proxy for TIR emission in the IRRC in the bottom row of Fig.~\ref{fig:qplot}. We see that most of the faint galaxies lie above the relationship determined for SFGs in \citet{Molnar2021}, however some of these lie in the extrapolated regions (indicated by the dashed lines) of the $q_{\rm TIR}$ relationships due to their low luminosities. NGC 1385 returns the expected $q_{\rm TIR}$ value in both plots and has a relatively high SFR compared to the rest of the galaxies within the Eridanus supergroup, suggesting that its possible radio continuum sensitivity limits have biased the $q_{\rm TIR}$ luminosity relationship to flatter slopes; however, our study is limited by the small sample size of seven SFGs. Future radio studies will be useful for constraining our results of the IRRC at the low luminosity end.

\begin{table*}[hbt!]
\centering
\begin{threeparttable}
\caption{Derived {\it WISE} quantities.}\label{WISEderived}
\begin{tabular}{c|c|c|c|c|c|c}
    \toprule
	Common ID & $W1-W2$ & $W2-W3$ & $L_{W3PAH}$ & $L_{\rm TIR}$ & $q_{12}$ & $q_{\rm TIR}$ \\
	 & (mag) & (mag) & ($log_{10}(L_{\odot})$) & ($log_{10}(L_{\odot})$) &  &  \\
	\midrule
	IC 1952 & $0.09\pm0.03$ & $3.03\pm0.04$ & $8.43\pm0.04$ & $9.7\pm0.3$ & $2.73\pm0.13$ & $4.0\pm0.5$ \\
	ESO 548-G036 & $0.20\pm0.03$ & $3.74\pm0.04$ & $8.56\pm0.04$ & $9.8\pm0.3$ & $1.94\pm0.06$ & $3.2\pm0.4$ \\
	IC 1953 & $0.09\pm0.03$ & $3.52\pm0.05$ & $8.85\pm0.04$ & $10.1\pm0.4$ & $2.30\pm0.09$ & $3.5\pm0.4$ \\
	NGC 1359 & $0.06\pm0.04$ & $3.21\pm0.05$ & $8.25\pm0.05$ & $9.5\pm0.3$ & $1.26\pm0.08$ & $2.6\pm0.4$ \\
	NGC 1367 & $0.03\pm0.04$ & $1.19\pm0.11$ & $8.18\pm0.11$ & $9.5\pm0.4$ & $1.34\pm0.07$ & $2.6\pm0.5$ \\
	NGC 1385 & $0.23\pm0.03$ & $3.87\pm0.04$ & $9.45\pm0.04$ & $10.6\pm0.4$ & $1.58\pm0.06$ & $2.7\pm0.4$ \\
	NGC 1415 & $0.10\pm0.03$ & $2.86\pm0.04$ & $8.94\pm0.03$ & $10.2\pm0.3$ & $1.86\pm0.06$ & $3.1\pm0.4$ \\
	NGC 1422 & $0.05\pm0.03$ & $2.69\pm0.04$ & $7.94\pm0.04$ & $9.3\pm0.3$ & $2.28\pm0.09$ & $3.6\pm0.4$ \\
    \bottomrule
\end{tabular}
\begin{tablenotes}[para]
	\item[]Note: Column (1): Galaxy common ID. Column (2): $W1-W2$ magnitude. Column (3): $W2-W3$ magnitude. Column (4): Total W3PAH luminosity. Column (5): Total estimated TIR luminosity. Column (6): global $q_{12}$ value. Column (7): global estimated $q_{\rm TIR}$ value.
\end{tablenotes}
\end{threeparttable}
\end{table*}

\subsection{Resolved IRRC properties of the Eridanus Supergroup}
\label{sec:resolvedirrc}

In this section we examine the resolved radio continuum and IRRC properties of the Eridanus galaxies, which are known to be experiencing environmental pre-processing, using the WALLABY continuum and {\it WISE} W3PAH maps. We build upon the analysis from the previous WALLABY pre-pilot survey papers to discuss the radio properties of these specific galaxies in the context of their HI distribution, HI deficiency, SFRs and disk stabilities \citep[][]{For2021, Murugeshan2021}, and tidal interaction strength \citep[][]{Wangs2022} to see whether we find any relationships between these properties and the resolved radio continuum emission. The criteria for HI deficiency in \citet{For2021} is defined as having less than half the measured HI mass compared to the expected HI mass estimated using the {\it R}-band absolute magnitude. In \citet{Murugeshan2021} HI deficiency is compared to the HI disk stability and hence there may be disagreement between these classifications. The values of stellar mass, HI masses and deficiency, and summed tidal parameter from \citet{For2021}, \citet{Murugeshan2021} and \citet{Wangs2022} are provided in Table~\ref{tab:HIprops} with a quantitave analysis available in previous WALLABY pre-pilot articles. Galaxy morphology classifications are from NED\footnote{https://ned.ipac.caltech.edu/}.

We see a number of trends which are consistent among this sample of galaxies. Firstly, radio continuum emission at this frequency is less extended than the W3PAH, optical and HI emission. Most of the radio emission appears to be produced in the core, due to higher SFRs, with certain galaxies exhibiting compact star-forming sites in their disks. \cite{Kennicutt1998} finds that for normal SF spirals the mean total SFR density can be as low as $\sim0.001\,M_{\odot}$yr$^{-1}$kpc$^{-2}$ with typical values of $\sim0.01\,M_{\odot}$yr$^{-1}$kpc$^{-2}$ derived using H$\alpha$ observations. The minimum 3$\sigma$ surface brightness limit we measure is 0.11\,mJy/beam, which corresponds to a SFR density of $0.02\,$M$_{\odot}$yr$^{-1}$kpc$^{-2}$ at 21\,Mpc suggesting we are limited by the surface brightness sensitivity to probe the low SFR density regions for these galaxies. W3PAH emission occurs throughout the stellar disk for all but NGC 1367. Across the sample most galaxies show little variation in the IRRC in the central regions with $q_{12}$ values consistent with SF ($q_{\rm 12} \sim 1.20 - 1.65$) and the radial extent being limited by the radio continuum sensitivity. The HI extent of these galaxies is larger than the radio continuum, IR, and optical extent and shows disturbed asymmetric morphology for all galaxies in the sample. A majority of the galaxies have $q_{12}$ maps with a lower median than their global $q_{\rm 12}$, determined in the previous section, due to the exclusion of extended IR flux from the final $q_{\rm 12}$ maps as a result of the more compact radio emission. Background radio sources and AGN activity are easily detected using the $q_{12}$ maps due to their significant localised radio excess causing them to be easily distinguishable from SF based IRRC measurements. This result may provide another method for AGN and background source detection usable for small, high-resolution samples of individual galaxies.

\subsubsection{IC 1952}
IC 1952 is loosely classified as a highly inclined S-shaped barred spiral galaxy. Due to a contamination of the {\it R}-band image by a foreground star, measurement of the expected HI mass is not possible hence \cite{For2021} could not determine whether IC 1952 is HI deficient or not; however, \citet{Murugeshan2021} do determine IC 1952 to be HI deficient as compared to its integrated HI disk stability. There is evidence of extraplanar gas with plumes of HI extending out of the disk as well as a slight extension towards the south-east. \citet{Wangs2022} determine this galaxy to be weakly perturbed by tidal interactions.

Due to the high angular resolution of the WALLABY continuum observations, we can easily distinguish the compact background radio source that lies behind IC 1952. We also observe localised SF occurring within the disk of IC 1952. The W3PAH emission occurs throughout the disk tracing the stellar and higher density HI emission. The ring of W3PAH emission to the south-east is caused by a foreground star. The $q_{\rm 12}$ map for this galaxy shows the star-forming regions as well as the radio excess caused by the background source. Combining all these observations we have no visible evidence for environmental effects outside of the HI distortion.

\begin{figure*}[hbt!]
	\includegraphics[width=\textwidth]{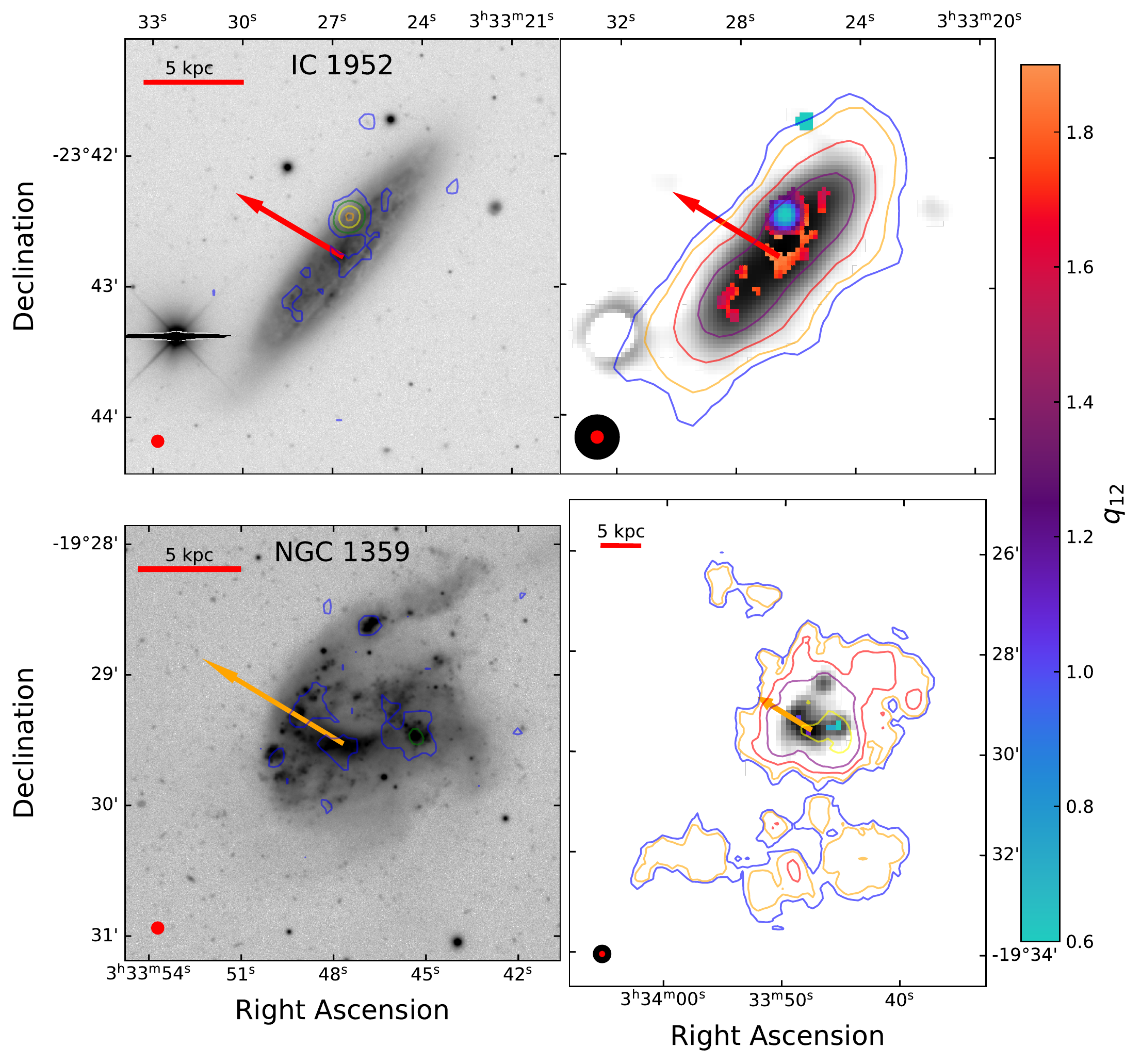}
    \centering
    \caption{The left column shows the 1.37\,GHz WC contours overlaid on the log-scaled SDSS {\it g}-band image for each galaxy, with contour levels of $3\sigma, 10\sigma, 25\sigma, 50\sigma, 100\sigma$ in blue, green, yellow orange and red respectively. The right column shows the $q_{\rm 12}$ map where both the radio and W3PAH emission are above $3\sigma$ with colours ranging from 0.6 (cyan - radio excess) to 1.9 (orange - radio deficient) overlaid on the {\it WISE} W3PAH image which is log-scaled between $3\sigma$ and its maximum value. The HI contours from \cite{For2021} are overlaid with contour levels of $1, 2, 5, 10, 20 \times10^{20}\,cm^{-2}$ in blue, orange, red, purple and yellow respectively. The red scale bar in the top left indicates 5\,kpc. The red and black filled circles in the bottom left indicate the beam size of WC/W3PAH (8$''$) and WALLABY HI (30$''$) respectively. The red arrow indicates the distance and direction to the Eridanus group central galaxy, NGC 1395, and is comparable between each galaxy (not scaled to the background image across galaxies). The orange arrow indicates the direction to NGC 1407, the central galaxy in the NGC 1407 group.}
    \label{fig:Murphy1}
\end{figure*}

\subsubsection{NGC 1359}
NGC 1359 is classified as a Magellanic S-shaped barred spiral galaxy in the coalescence stage of a minor merger \citep{Omar2005}. \citet{Murugeshan2021} measures NGC 1359 to have enhanced specific SFR (sSFR) while its atomic gas fraction is higher than expected making it HI-normal for its measured integrated disk stability as compared to other Eridanus group members. This enhanced atomic gas fraction and sSFR is likely a result of the accretion of fresh gas during the merger which leads to increased SF. \citet{Wangs2022} find this galaxy is the least affected by tidal interactions and has the highest size ratio (between HI and stellar disk) in our sample; however, as the tidal interaction measurement only takes into account perturbations from galaxies outside this system, it does not take into account the tidal impact of the past minor merger.

This galaxy exists at the outer edge of the WALLABY Eridanus field in an edge beam containing no footprint overlap. As such the sensitivity of the WALLABY observations is lower. The radio continuum emission does trace the starburst sites extremely accurately. The W3PAH emission traces the {\it g}-band emission however it does not extend beyond the stellar disk unlike most of the galaxies in this sample. There is a significant flux deficit in measured W3PAH emission suggesting that dust may have been removed and PAHs destroyed during the merger and subsequent starbursts; however, we see no evidence of extended W3PAH emission tracing the HI gas as we do in NGC 1367 or NGC 1415.

Due to a significant W3PAH flux deficit, we find very low $q_{12}$ values indicative of AGN or background radio sources; however, due to their distribution and coincidence with starburst sites, we rule out this possibility. We do not observe evidence of environmental effects in the $q_{\rm 12}$ map because of the decreased sensitivity of WC observations in this region of the field. We do however see that NGC 1359 has an extremely disturbed stellar and HI morphology with a massive debris field to the south as a result of the minor merger that this galaxy has undergone.

\subsubsection{IC 1953}
IC 1953 is classified as a face-on barred spiral galaxy with a mixed inner ring. It is determined to be HI deficient in comparison to both its expected HI mass and integrated HI disk stability. As shown in Fig.~\ref{fig:Murphy1} there is an HI extension towards ESO 548-G036 that lies to the south-west. This is suggestive of interaction between these galaxies, since \citet{Wangs2022} measure this galaxy is only weakly perturbed by the interaction.

The radio continuum emission is primarily produced in the core of the galaxy but we do also see localised emission in the arms of the galaxy both of which coincide with increased HI density suggesting SF is occurring in these areas. It is likely that the extended radio emission is resolved out as we only measure about half the total NVSS flux. \citet{Omar2005} find that there is diffuse emission throughout the disk and that there is evidence of a nuclear starburst \citep{Roussel2003}. We see that the W3PAH emission occurs throughout the stellar disk with localised increases in the spiral arms corresponding to the star forming areas observed in the radio emission. An extension towards the south-west is also seen in the W3PAH emission in the direction of ESO 548-G036 tracing the HI extension.

The $q_{\rm 12}$ map for this galaxy does not show any evidence of environmental effects or asymmetries. The $q_{\rm 12}$ map does however provide evidence that the central emission is unlikely to be caused by AGN activity because the values are roughly consistent with those related to SF in the arms and not significantly radio excess such as in NGC 1367. Overall, evidence for environmental effects are only seen in the HI and W3PAH maps for this galaxy.

\begin{figure*}[hbt!]
	\includegraphics[width=\textwidth]{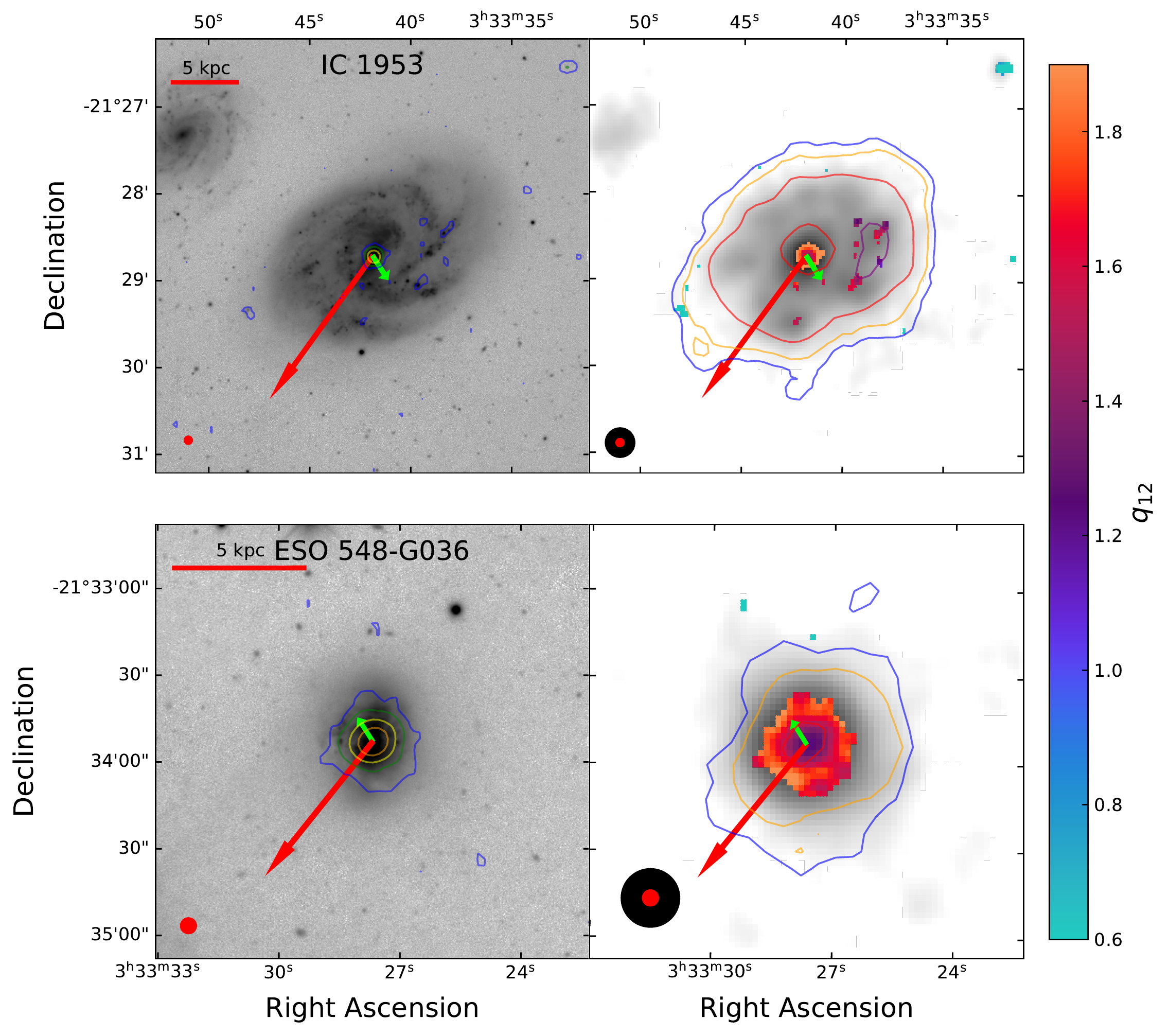}
    \centering
    \caption{Continued, with the addition that the lime green arrows indicate the distance and direction between IC 1953 and ESO 548-G036 and are scaled by a factor of 3 with respect to the red arrow for visibility.}
\end{figure*}

\subsubsection{ESO 548-G036}
ESO 548-G036 is classified as a face-on spiral galaxy and is measured to be HI deficient with respect to its expected HI mass (see appendix Table.~\ref{tab:HIprops}). ESO 548-G036 lies south-west of IC 1953 and shows a disturbed HI morphology with a HI cloud detected north of the galaxy. This galaxy is also classified as weakly perturbed by tidal interactions in \citet{Wangs2022}.

ESO 548-G036 has radio emission throughout the stellar disk suggesting SF is occurring throughout the galaxy. Flux does not appear to be resolved out for this galaxy due to its small angular size giving us comparable total flux values to NVSS. The W3PAH emission extends beyond the stellar disk and roughly coincides with the HI distribution. 

The $q_{\rm 12}$ map has values consistent with SF and does not show a statistically significant gradient across the disk thus providing no additional evidence for interaction with IC 1953.

\subsubsection{NGC 1367}

\begin{figure*}[hbt!]
	\includegraphics[width=\textwidth]{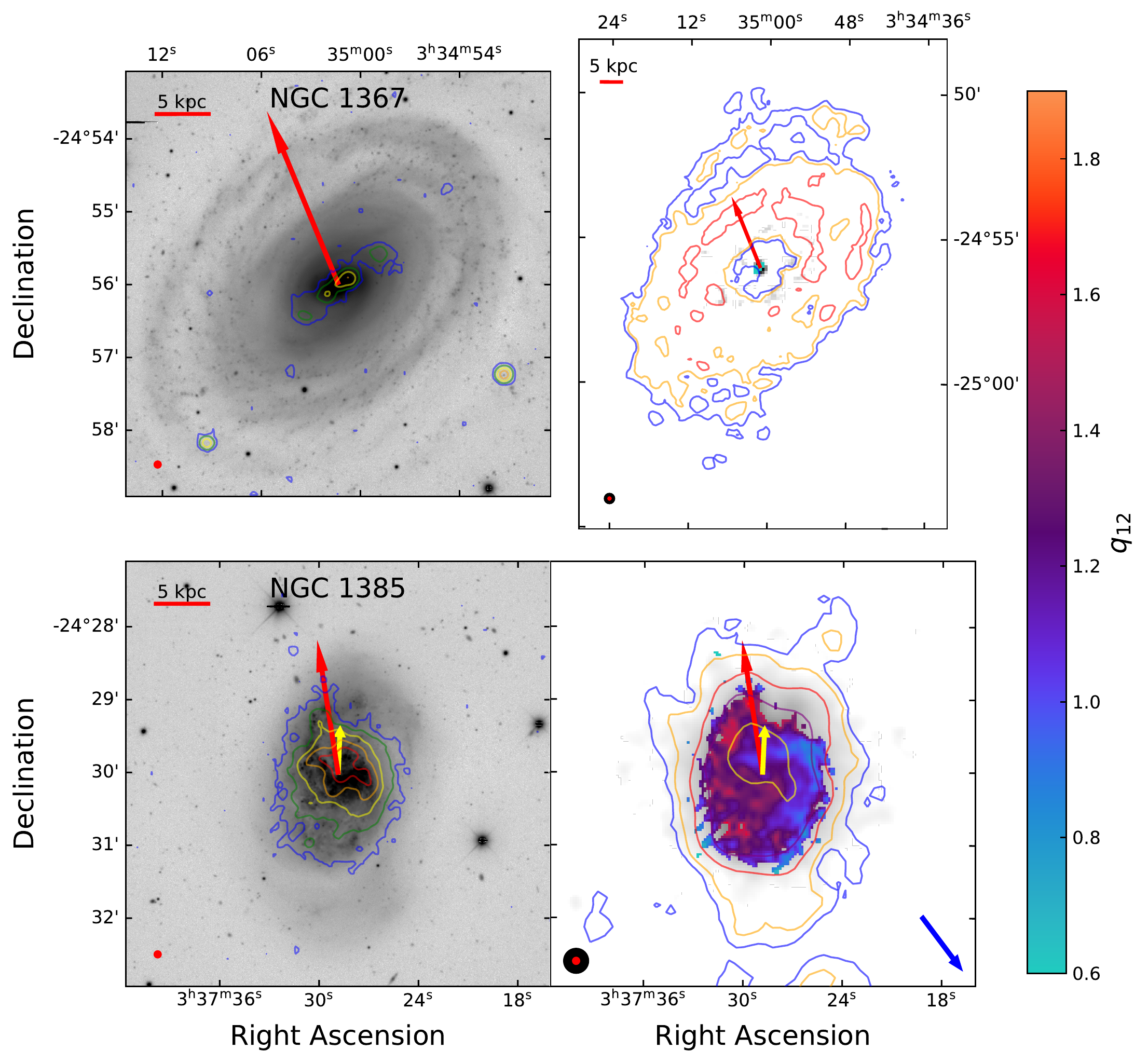}
    \centering
    \caption{Continued, with the addition that the yellow arrow indicates the distance to Dark cloud 2 \citep{Wong2021} and is directly comparable to the red arrow. The blue arrow indicates the likely direction of CR electron streaming or increasing radio continuum emission.}
\end{figure*}

NGC 1367 is classified as a slightly inclined mixed-bar spiral galaxy with a mixed inner ring and star forming outer ring. The HI deficiency parameter with respect to its expected HI mass for this galaxy in not measured due to {\it R}-band contamination as well as missing a large amount of extended HI flux. \citet{Murugeshan2021} however do find this galaxy to be slightly HI deficient with respect to its integrated HI disk stability. NGC 1367 is a large galaxy with an extremely extended HI diameter of 22.3 arcmin as observed by the $9.1'$ observations by the Green Bank Telescope \citep[][]{Sorgho2019}, however WALLABY only recovers an HI diameter of $9.9'$ and $83\%$ of the total HI mass.

We observe double radio lobes and jets in this galaxy. The radio jets of this galaxy have half the extent of the inner stellar disk, lie slightly offset from the bar, and coincide with the HI hole. This structure suggests that NGC 1367 may be in the same class of galaxies as other nearby spirals with small radio jets including Circinus \citep{Harnett1990, Koribalski2008} and NGC 3079 \citep{Wong2016}. \citet{Omar2005} observe NGC 1367 has a nuclear radio source coincident with the $H\alpha$ nuclear source observed by \citet{Hameed1999}; however, no $H\alpha$ is seen coincident with the extended jet structure outside weak diffuse emission in the bulge. We also do not observe significant amounts of radio or W3PAH emission in the bulge suggesting that this galaxy is not strongly star-forming in this region. NGC 1367 displays UV and $H\alpha$ emission in an outer ring which is faintly visible in the {\it g}-band image and roughly coincides with the W3PAH emission. This observation suggests that SF is occurring in this region however either due to surface brightness sensitivity limits or resolving out this extended flux we do not see it in the WC. NGC 1367 also appears in the Chandra X-ray survey and contains a nuclear source with surrounding diffuse emission which is analysed in detail \citep[see NGC 1371;][]{Hughes2007} with the conclusion that NGC 1367 contains a low-luminosity AGN similar to other nearby spiral galaxies.

The $q_{\rm 12}$ map for this galaxy clearly shows the radio excess due to the radio AGN in the central region and provides a good indicator for the separation of AGN and SF activity based on its value. Overall due to the existence of an AGN within this galaxy and its low SF, environmental effects are not discernible.

\subsubsection{NGC 1385}
NGC 1385 is classified as an almost face-on \citep[$i = 44^{\circ}, P.A = 181^{\circ}$,][]{Leroy2021} barred spiral galaxy with S-shaped bars and is the highest SFR galaxy in our sample (see Table.~\ref{tab:SFRtab}). NGC 1385 is located in the south-west of the Eridanus group, is $0.6^{\circ}$ south of {\it WALLABY J033723--235753} (C2 in \citet{Wong2021}), an HI cloud with no detected stellar component. \citet{For2021} measure this galaxy to be HI deficient as compared to its expected HI mass as estimated by its stellar mass. The HI morphology of NGC 1385 is extremely distorted with clear compression and extensions in the north-east and south-west respectively. This galaxy also exhibits a southern tidal debris field (which is partially cut off in Fig~\ref{fig:Murphy1}) with no optical counterpart suggesting HI gas has been displaced likely due to tidal interactions with its nearby companion {\it WALLABY J033723--235753}. Galaxies that are currently interacting and accreting HI gas from their dwarf companions can have enhanced SFR \citep[][]{Kennicut1987, Ellison2011, Murugeshan2020}. \citet{Murugeshan2021} measures NGC 1385 to have enhanced sSFR while its atomic gas fraction is higher than expected for the measured integrated disk stability as compared to other Eridanus group members. This means that compared to other galaxies within the Eridanus group this galaxy is one of two that is actually HI-normal for their integrated disk stability, likely due to the accretion of HI from its interacting companion. This accretion of fresh gas for NGC 1385 leads to increased HI mass and SF which is consistent with the findings of \citet{Ellison2018} and \citet{Murugeshan2020}. This galaxy is found to have extremely weak tidal interaction parameters due to its mass and location within the Eridanus subgroup \citep{Wangs2022}, however this does not take into account the interaction with {\it WALLABY J033723--235753} due to the lack of a stellar component.

NGC 1385's disk is almost face on and the SFR is relatively high so this is a good candidate for examining whether environmental effects (detected in the HI) are also observed in the radio continuum. We see strong radio continuum emission throughout the stellar disk except in the northernmost region likely due faint extended emission being below our surface brightness sensitivity limit. We observe radio emission in the south-western edge of NGC 1385 that extends slightly beyond the stellar disk. The W3PAH emission does not show any clear signs of disturbance and is observed throughout the galaxy.

The $q_{\rm 12}$ map of NGC 1385 shows slightly lower values due to enhanced radio emission possibly caused by interactions with the environment. We can see a clear gradient of decreasing $q_{\rm 12}$ moving from the north-east the the south-west of the galaxy (as shown by the blue arrow in Fig.~\ref{fig:Murphy1}. This roughly coincides with the direction of the group centre and {\it WALLABY J033723--235753} suggesting that ram-pressure and/or tidal interactions may be leading to CR electron reacceleration and streaming which is forming a small synchrotron tail. This trend agrees with the findings of that from \citet{Murphy2009} and \citet{Vollmer2020}, which suggests that CR electron streaming results in a redistribution of CR electrons and hence radio continuum emission. This bulk motion is not reflected in the ISM emission measured by W3PAH due to different scale lengths of the mean free path of ionising UV photons and CR electrons. The direction of the gradient in the IRRC also coincides with the disturbance in HI morphology possibly hinting at their common origin. 

\subsubsection{NGC 1415}
NGC 1415 is classified as a moderately inclined spiral galaxy with S-shaped mixed bars and an outer ring. It is located in the central region of the Eridanus group and is the closest galaxy to the group centre in our sample. NGC 1415 has been measured to be HI deficient as compared to its expected HI mass and integrated disk stability. \citet{Wangs2022} find NGC 1415 to be weakly perturbed by tidal interactions; however, it has the second highest summed tidal interaction strength for this sample.

The HI traces the stellar disk and outer ring whilst radio continuum emission is produced primarily in the central region of the galaxy. W3PAH emission occurs throughout and slightly beyond the stellar disk tracing the HI cloud. It is likely that we are missing the extended radio emission in the outer disk as we are missing flux compared to NVSS. Evidence of a nuclear geyser (bipolar outflow) is apparent in the $H\alpha$ and measured kinematics of NGC 1415, which is suggested to be caused by AGN \cite{Garcia2019}; however, we do not identify any jet structure within NGC 1415. This galaxy shows a truncated star forming disk compared to the entire stellar disk. Truncated star forming disks are seen in Virgo spirals that are thought to have experienced ram-pressure stripping \citet{Koopmann2004, Vollmer2012}.

The $q_{\rm 12}$ map shows no significant variation within the galaxy and mostly traces the SF occurring in the central region. A slight but statistically insignificant gradient can be seen with higher radio emission observed to the south-west of the galactic centre and in the direction of the group centre. However, we do not see any corresponding distortion in the HI morphology in this direction. We also see a background AGN south-west of NGC 1415. Combining these features we do not have conclusive evidence of ram pressure stripping or tidal interactions.

\begin{figure*}[hbt!]
	\includegraphics[width=\textwidth]{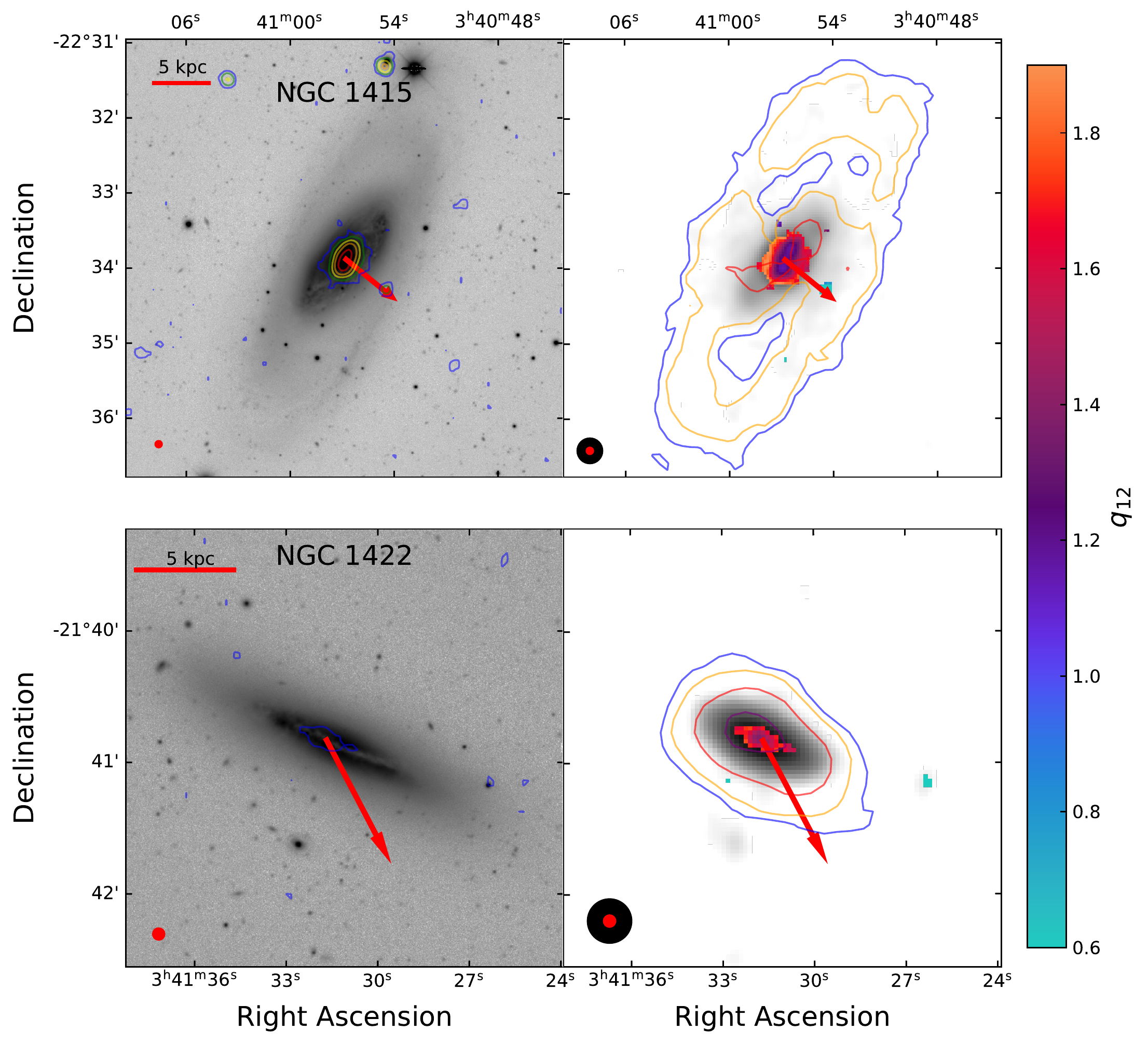}
    \centering
    \caption{Continued.}
\end{figure*}

\subsubsection{NGC 1422}
NGC 1422 is classified as an edge-on barred spiral galaxy and is the lowest stellar mass and faintest galaxy in our sample. It is located in the north-eastern region of the Eridanus group and is measured to be HI deficient. The HI disk for NGC 1422 is truncated compared to the stellar disk, and extends beyond the galactic plane suggesting ram-pressure stripping is occurring within this galaxy \citep{Murugeshan2021}. \citet{Murugeshan2021} also find this galaxy to be HI deficient with respect to its HI disk stability parameter. \citet{Wangs2022} find this galaxy to be strongly perturbed by tidal interactions and is the only galaxy in our sample with this classification.

NGC 1422 has the lowest measured SFR and no AGN activity leading to extremely low radio continuum emission that only shows up at the $3 \sigma$ detection limit in the core of the galaxy and is undetected in NVSS. Similar to the HI distribution, we see a truncation of the W3PAH emission at the edge of the stellar disk and an extension beyond the galactic plane, which suggests the heated dust and ISM is also being affected by ram-pressure stripping.

The $q_{\rm 12}$ map for this galaxy does not provide evidence for environmental effects due to the extent of radio emission being limited to the core of the galaxy, however we can likely rule out AGN related radio emission due to the values being consistent with SF activity.

\section{Discussion}

\subsection{Global Continuum Properties}

We find (see Table. \ref{tab:WCDerived}) that the global spectral indices of the eight Eridanus supergroup members (after removal of outliers) are generally steeper than expected of normal SFGs. A number of effects could be at play which are causing this steepness. Firstly the previously mentioned flux scale issues with RACS and WC will act to steepen the measured spectral index for these galaxies however we can still compare them to each other. It has been shown that SF flattens the spectral indices and that compact starbursts have a flatter spectral index primarily thought to be caused by free-free absorption \citep{Murphy2013, An2021}. As the Eridanus supergroup galaxies generally have lower levels of SF this steeper spectral index may be somewhat expected. Within our sample we generally observe the higher SFR galaxies to have a flatter spectral index. We also see that the spectral index of NGC 1385 and NGC 1359, the two systems with clear signs of past interactions in their HI morphology, have spectral indices which agree with the results in \citet{Murphy2009}, \citet{Murphy2013} and \citet{Donevski2015}. NGC 1385 has a slightly steeper spectral index and is thought to be currently undergoing tidal interactions which act to steepen the spectral index, whilst NGC 1359 is post merger and contains compact starbursts which will flatten the spectral index.

The SFRs for our sample of Eridanus galaxies are largely consistent with  previously derived luminosity-SFR relationships \citep{Cluver2014, Davies2017, Wang2017, Molnar2021} (Fig. \ref{fig:SFR}) even whilst probing their lower limits. Significant outliers can be explained by the underestimation of radio flux (IC 1952, IC1953) or possible PAH destruction (NGC 1367, NGC 1359). It also supports that the radio-SFR provides an accurate estimate of the current SFR. \citet{Moorman2016, Murugeshan2020} have shown that SFRs of galaxies undergoing interactions can be modified and that the sSFR of these Eridanus group galaxies are generally lower (except for NGC 1385 and NGC 1359 due to the accretion of fresh HI gas from their companions) than those from the field \citet{Murugeshan2021}.

We find a steeper dependence on W3PAH luminosity than radio luminosity in contrast to the results for $q_{\rm TIR}$ \citep{Moric2010, Molnar2021}. The estimated TIR luminosities are significantly higher than expected resulting in discrepancies in $q_{\rm TIR}$ as compared to \citet{Molnar2021}. We suggest that the scaling relationship between W3PAH and TIR luminosities determined for SINGS/KINGFISH galaxies (Eq.~\ref{eq:LTIR}) may not be appropriate for our sample due to their low sSFRs which may be less severely impacted by PAH destruction. This discrepancy may be caused by the destruction of PAHs in starbursts \citep{draine07, Cluver2017} which will impact galaxies with higher SFRs more severely, causing the observed W3PAH emission to increase more slowly than TIR emission and hence biasing $q_{12}$ towards lower values with increasing W3PAH luminosity. 

NGC 1359 shows a significant deviation compared to the other SFGs due to its low W3PAH emission, which is possibly due to merger-induced PAH destruction \citep{Appleton2006, micelotta10, murata17} and is discussed in more detail in Section 6.3. When compared to the luminosity trends observed in \citet{Molnar2021} we find that as we move towards lower luminosities the galaxies begin to deviate from their expected properties. This is likely because these galaxies have low radio luminosities (partially due to missing flux problems) and probe the lower limits of current radio surveys and so are not well sampled. It may also hint at the luminosity limit of the expected breakdown of the IRRC due to UV and optical photons not being fully reprocessed by dust \citep{Bell2003, Lacki2010}, however this would decrease observed $q_{\rm 12}$ values which are contradictory to our findings. Previous studies by \citet{Miller2001} and \citet{Reddy2004} have found the global $q$-value to be lower in cluster galaxies compared to those in the field which is attributed to thermal compression and ram pressure based enhancement of the synchrotron emission. Due to our approximate values for the TIR luminosity and radio flux deficiency we do not see this. It is also worth mentioning that these galaxies exist in a group environment which is less dense and undergoing pre-processing before becoming a supergroup and hence have not experienced the strong interactions that occur in cluster environments. Overall we find that while W3PAH emission can be used to measure the IRRC globally for galaxies because of the tight correlation between W3PAH and TIR luminosities \citep{Cluver2014, Cluver2017}, the impact of PAH destruction and even initial PAH concentration due to differing metallicities has to be taken into account. Future studies on the relationship between W3PAH emission and $q_{12}$ will further verify our conclusions.

\subsection{Resolved Continuum Properties}

We find that the $q_{\rm 12}$ values in star-forming regions are largely consistent with each other across our sample providing evidence that the IRRC derived using W3PAH is a result of SF processes. We also find that $q_{\rm 12}$ is lower in the centre of each galaxy. This suggests that, unlike $q_{\rm TIR}$ which is enhanced in star-forming regions \citep{Tabatabaei2013, Vollmer2020}, the interstellar radiation field becomes hard in these areas destroying the PAHs and lowering the W3PAH emission (and hence $q_{\rm 12}$) which is not entirely offset by the increase in $W3$ continuum emission produced by the large warm grains. This is possibly caused by the existence of radio-quiet, weak AGN, but, due to the lack of X-ray and spectroscopic data in this field, we are unable to assess this possibility for each galaxy.

The {\it WISE} color-color diagram suggests that these galaxies are unlikely to harbor powerful AGN (see Fig.~\ref{fig:WISEcmd}) as they exist well below the $W1-W2 \geq 0.5$ region where AGN sources typically exist \citep{Jarrett2013}. Following the findings of \citet{For2021}, \citet{Murugeshan2021} and \citet{Wangs2022}, we conclude that the environmental pre-processing for these galaxies has removed HI gas and suppressed the collapse of HI gas into molecular clouds and hence decreased the sSFR for these galaxies without a significant enhancement due to tidal interactions. A major finding for these low SFR galaxies is that the resolved $q_{\rm 12}$ maps also allow us to easily distinguish background radio sources (IC 1952) and AGN activity due to a significant decrease in measured $q_{\rm 12}$ value. 

In the case of NGC 1359 however the $q_{\rm 12}$ map suggests that due to the merger and enhanced sSFR, PAH destruction likely plays a significant role in lowering the measured $q_{\rm 12}$ value. It is likely then that this deficit in W3PAH emission is caused by PAH destruction either by shocks during the merger or the subsequent starburst activity. Indeed it has been shown that mergers especially during the late stages can destroy PAH's via large scale shocks \citep{murata17}, both low and high speed, which can cause thermal and inertial sputtering \citep{micelotta10}. Mergers with high collision speeds ($>500\,$km/s) such as those in Stephan's quintet are also known to destroy PAHs via high speed, highly dissociative shocks \citep{Hollenbach1980, Hollenbach1989, Appleton2006}. 

NGC 1367 is interesting due to its AGN jets with distinct radio lobes, very little W3PAH emission and a $> 10'$ HI extent. We do not observe evidence of any environmental effects in the $q_{12}$ map. The bulge does not appear to be forming stars at a significant rate and has very little ISM emission in the W3PAH band suggesting either a lack of UV photons being produced or a lack of hot dust and PAHs. We see the strongest W3PAH emission coincident with the location of the AGN where dust heating and PAH destruction would be most prevalent. The absent W3PAH emission combined with the lack of UV and $H\alpha$ emission in the bulge suggests that it is likely very little SF is occurring, with most current SF occurring in the outer spiral arms as traced by UV observations. The HI hole in the central region of this galaxy is coincident with the jet structure and suggest the possibility that the AGN activity may be related to the lack of HI gas in this central region. Jet-ISM interactions have been posited to both suppress (via gas ejection, suppressing halo gas accretion or induced turbulence) and enhance (via shock compression) SF depending on a large number of factors \citep[see][]{Mukherjee2018, Mandal2021, Mukherjee2021, Girdhar2022}. However, an in depth analysis of jet-ISM interactions is beyond the scope of this paper.

NGC 1385, the galaxy with the highest SFR in the Eridanus group, is understood to be undergoing tidal interactions with a nearby HI cloud. We see slight evidence of a small synchrotron tail in the south of NGC 1385 which suggests CR electron streaming is occurring due to tidal interactions or interactions between the ISM and intragroup medium (IGrM) wind. This IGrM wind is theorised to cause shocks which travel quickly through the ISM transporting and reaccelerating CR electrons leading to synchrotron tails and radio deficit regions at the leading edge of the galaxy \citep{Murphy2009}. \citet{Murphy2009} and \citet{Vollmer2020} show that this electron streaming as a result of environmental effects leads to significant differences in the measured vs expected radio emission across the galactic disk. These significant differences are not typically observed in secularly evolving galaxies \citep{Murphy2008}, however there can be a number of factors at play which affect the synchrotron emission within individual galaxies such as magnetic field strength, SFR, radiation fields and neutral gas \citep{Tabatabaei2013}.
 
The $q_{\rm 12}$ map shows a clear increase (radio deficit) in the north-east of NGC 1385, exterior to the spiral arm, indicating either a lower CR electron density, weakening of the turbulent magnetic fields \citep{Murphy2009, Vollmer2020} or conversely an excess of W3PAH emission. Due to a lack of polarisation observations we can not measure the strength of the magnetic fields within this galaxy and are unable rule out synchrotron enhancement by compressed magnetic fields. An excess of W3PAH emission also seems unlikely as it has been shown that W3PAH emission appears to have a weaker correlation with SFR, particularly as the SFR density reaches a critical point whereby the PAHs can be destroyed. Therefore we posit that IGrM-ISM and tidal interactions which transport the CR electrons towards the south-west is likely to be the primary cause of the radio deficit in this region. The south-west of the galaxy then shows decreased $q_{\rm 12}$ values indicating a radio excess. We also do not observe as prominent of a star forming spiral arm further providing evidence that the CR electron density here is enhanced rather than SF based magnetic field enhancement \citep{Tabatabaei2013} or PAH destruction causing a W3PAH deficit.  Tidal interactions have also been shown to cause tidal CR electron emission to dominate the global CR electron emission and hence decrease $q_{\rm 12}$ locally. The $q_{\rm 12}$ gradient almost perfectly aligns with the disturbances in HI morphology whereby we see compression in the HI in the northern edge of NGC 1385 and extension with a tidal debris field in the south. These features in turn all align with the direction to the group centre as well as the interacting companion suggesting that they are the cause of these variations in $q_{\rm 12}$ and HI morphology. 

\section{Summary and Conclusions}
In this paper we present the radio continuum catalogue at 1.37\,GHz from pre-pilot WALLABY observations of the Eridanus field using the ASKAP telescope.
\begin{enumerate}
    \item We demonstrate that ASKAP produces excellent continuum image quality with a sensitivity down to 0.1\,mJy and mean RMS of 0.05\,mJy/beam at an angular resolution of $6.1\,''\times7.9\,''$. We detect 9461 components in the field and find that the WC flux density measured using {\tt ProFound} is consistent within the scatter with NVSS at 1.4\,GHz.
    \item Our flux offset and spectral index analysis suggests that WC slightly underestimates the flux by $\sim5\%$ for isolated unresolved sources and that the RACS catalogue slightly overestimates flux despite both being accurate within their estimated scatter.
    \item These pre-pilot Eridanus observations enable the study of the radio-SFR and IRRC correlation down to lower SFRs and galaxies with lower stellar masses. We find the 1.37 GHz radio continuum provides an accurate estimate of the total SFRs within our sample of eight galaxies (aside from NGC 1367 due to AGN contamination) compared to those determined using multi-wavelength methods and W3PAH emission. We also probe the global IRRC for the Eridanus group galaxies using W3PAH emission and find that the general trends from literature using FIR or TIR emission are reproduced for SFGs.
    \item The sensitivity limit of WC observations is slightly too high to detect the faint diffuse radio continuum emission due to SF throughout the disks for such a low stellar mass and SFR sample. This prevents studies of the environmental effects and their impact on the radio continuum emission and the IRRC for most of our sample. However we do find that the resolved IRRC can be used to easily identify background radio sources and discriminate between SF and AGN based radio continuum emission despite the sensitivity limitations.
    \item The impact of environmental interactions are not as evident in the radio continuum morphology as it primarily traces local star forming regions within the galactic disks which are less susceptible than outer regions to weak interactions with the environment. 
    \item We confirm the existence of a double lobed radio jet within NGC 1367. The radio jet is coincident with holes in the HI emission. SF appears to be suppressed in the bulge of this galaxy as supported by the lack of W3PAH and UV emission in this region.
    \item NGC 1359 is in the coalescence stage of a merger and has a significant deficit of W3PAH emission. We suggest that merger induced shocks may have destroyed the PAHs in this galaxy however MIR spectroscopy is needed to confirm whether this is the case.
    \item NGC 1385 shows clear signs of environmental pre-processing in its radio continuum and resolved IRRC maps. We observe a small synchrotron tail which coincides with the direction of the HI tidal debris field. The resolved IRRC shows a clear gradient from north-east to south-west suggesting CR electron streaming is occurring in this direction. The resolved spectral index map between RACS and WC for this galaxy has a slight gradient in the same direction however poor angular resolution and flux scale errors limit the conclusions drawn from the resolved spectral index map.
\end{enumerate}

Overall we find that the radio continuum emission observations obtained during the WALLABY pre-pilot observations of the Eridanus group allow the resolved study of the SF properties of SFGs and AGNs. These SF properties can be compared to their HI properties and other ancillary multiwavelength observations to provide a clearer picture of the physical processes which modify the IRRC within galaxies. Our understanding of these processes will only improve with the completion of the full WALLABY survey and more sensitive radio surveys such as EMU. Thus by understanding the physical processes which cause deviations from the expected radio emission, spectral index and IRRC we will be better able to characterise galaxies in larger samples of sources.

\section*{Acknowledgements}

The Australian SKA Pathfinder is part of the Australia Telescope National Facility which is managed by CSIRO. Operation of ASKAP is funded by the Australian Government with support from the National Collaborative Research Infrastructure Strategy. ASKAP uses the resources of the Pawsey Supercomputing Centre. Establishment of ASKAP, the Murchison Radio-astronomy Observatory and the Pawsey Supercomputing Centre are initiatives of the Australian Government, with support from the Government of Western Australia and the Science and Industry Endowment Fund. We acknowledge the Wajarri Yamatji people as the traditional owners of the Observatory site. This publication makes use of data products from the Wide-field Infrared Survey Explorer, which is a joint project of the University of California, Los Angeles, and the Jet Propulsion Laboratory/California Institute of Technology, funded by the National Aeronautics and Space Administration. This research has made use of the NASA/IPAC Extragalactic Database (NED), which is operated by the Jet Propulsion Laboratory, California Institute of Technology, under contract with the National Aeronautics and Space Administration. Parts of this research were supported by the Australian Research Council Centre of Excellence for All Sky Astrophysics in 3 Dimensions (ASTRO 3D), through project number CE170100013. We thank the referee for their in-depth review and comments on this manuscript.

\section*{Data Availability}

The data underlying this article are available in the article and in the online supplementary material. The processed ASKAP data can be retrieved via CSIRO ASKAP Science Data Archive (CASDA) with a given scheduling block identification number. The mosaicked continuum map of the Eridanus field observations is available at the DOI: https://doi.org/10.25919/8ga8-0n09.

\bibliography{eridanus}

\appendix

\section{ASKAP Observation Details}

\begin{table*}[hbt!]
\centering
\begin{threeparttable}
\caption{ASKAP Observation Details.}\label{tab:Obstab}
\begin{tabularx}{0.98\textwidth}{ccccccccc} 
	\toprule
	UT date & Footprint & RA & Dec. & Calibrator SBID & Science SBID & Integration time & Bandwidth & Central frequency\\
	(yyyy-mm-dd) & & ($^{h}:^{m}:^{s}$) & ($^{\circ}:':''$) & & & (h) & (MHz) & (MHz)\\
	\midrule
	2019-03-13 & A & 3:39:30 & -22:30:00 & 8169 & 8168 & 5.8 & 288 & 1295.5\\
	2019-03-13 & B & 3:36:44.52 & -22:37:54.69 & 8169 & 8170 & 5.0 & 288 & 1295.5\\
	\bottomrule
\end{tabularx}
\begin{tablenotes}[para]
    \item[]Note: Column (1): universal time date. Column (2): footprint designation. Column (3): J2000 right ascension. Column (4): J2000 declination. Column (5-6): calibrator and science SBIDs. Column (7): total integration time. Column (8-9): bandwidth and central frequency.
\end{tablenotes}
\end{threeparttable}
\end{table*}

\section{HI Properties of Eridanus Supergroup Members}

\begin{table*}[hbt!]
\centering
\begin{threeparttable}
\caption{HI properties of Eridanus supergroup members}\label{tab:HIprops}
\begin{tabular}{c|c|c|c|c|c|c}
    \toprule
	Common ID & Log($M_{*}^{M21}$) & Log($M_{HI}^{For21}$) & $DEF_{HI}^{For21}$ & Log($M_{HI}^{M21}$) & $\Delta\,f_{qHI}^{M21}$ & $S_{SUM}$ \\
     & ($M_{\odot}$) & ($M_{\odot}$) &  & ($M_{\odot}$) &  & \\
    \midrule
	IC 1952 & $9.65\pm0.07$ & $8.72\pm0.20$ & $^b$ & $8.80\pm0.21$ & 0.35 &  $-2.35\pm0.11$ \\
	ESO 548-G036 & $9.43\pm0.23^{a}$ & $8.01\pm0.23$ & $0.95$ & $^c$ & $^c$ & $-1.87\pm0.38$ \\
	IC 1953 & $9.93\pm0.06$ & $8.97\pm0.20$ & $0.46$ & $ 9.07\pm0.02$ & $0.21$ & $-1.99\pm0.23$ \\
	NGC 1359 & $9.42\pm0.06$ & $9.41\pm0.20$ & $^b$ & $9.56\pm0.20$ & $-0.07$ & $-2.62\pm0.15$ \\
	NGC 1367 & $10.30\pm0.06$ & $9.76\pm0.20$ & $^b$ & $9.64\pm0.20$ & $0.10$ & $^d$ \\
	NGC 1385 & $10.08\pm0.06$ & $9.34\pm0.20$ & $0.34$ & $9.24\pm0.20$ & $-0.12$ & $-2.47\pm0.14$ \\
	NGC 1415 & $10.20\pm0.08$ & $8.93\pm0.20$ & $0.63$ & $8.87\pm0.21$ & $0.30$ &  $-1.78\pm0.12$ \\
	NGC 1422  & $9.22\pm0.09$ & $8.33\pm0.21$ & $0.66$ & $8.32\pm0.23$ & $0.38$ & $-1.61\pm0.08$ \\
    \bottomrule
\end{tabular}
\begin{tablenotes}[para]
	\item[]Note: Column (1): Galaxy common ID. Column (2): Stellar mass from \citet{Murugeshan2021}. Column (2): Stellar mass from \citet{Murugeshan2021}. Column (3): HI mass from \citet{For2021}. Column (4): HI deficiency parameter as compared to stellar mass from \citet{For2021}, galaxies above 0.3 are considered HI deficient. Column (5): HI mass from \citet{Murugeshan2021}. (6): HI deficiency parameter as compared to disk stability from \citet{Murugeshan2021}, galaxies above 0.2 are considered HI deficient. Column (7): Summed tidal parameter from \citet{Wangs2022} values less than -1.69 are considered to be weakly perturbed by tidal interactions.
    \item[a]Stellar mass from \citet{For2021} as this galaxy did not meet selection criteria in \citet{Murugeshan2021}.
    \item[b]Stellar mass is not determined in \citet{For2021} thus a HI deficiency is not measured.
    \item[c]This galaxy did not meet selection criteria in \citet{Murugeshan2021}.
    \item[d]This galaxy did not meet selection criteria in \citet{Wangs2022}.
\end{tablenotes}
\end{threeparttable}
\end{table*}

\section{WC Catalogue {\tt Profound} Input Parameters and Extended Catalogue Column Descriptions}
\label{sec:catdesc}

\begin{table}[hbt!]
\centering
\begin{threeparttable}
\caption{Modified ProFound input parameters.}\label{tab:inptab}
\begin{tabular}{cccccc}
    \toprule
	box & skycut & pixcut & tolerance & reltol & ext\\
    \midrule
	100 & 3 & 3 & 10 & 1 & 4\\
	\bottomrule
\end{tabular}
\begin{tablenotes}[para]
    \item[]Note: Descriptions for each input parameter can be found in \citet{Robotham2018} or at https://rdrr.io/cran/ProFound/man/profoundProFound.html.
\end{tablenotes}
\end{threeparttable}
\end{table}

We use a subset of the final column information generated by the {\tt ProFound} to generate the final simplified and extended WC catalogues. These columns provide information on: IDs; astrometry; flux densities; shape information and other important source information. The first five lines of the simplified WC source catalogue are presented in Table.~\ref{tab:catalogue} sorted by R.A which includes a subset of the columns described below.

\begin{itemize}
    \item {\tt Designation} - The designation of the source given in IAU convention JHHMMSS$\pm$DDMMSS with the prefix WC.
    \item {\tt segID} - The final segmentation ID number of the source.
    \item {\tt segID2} - The initial segmentation ID number of the source before performing $\sigma$ cuts.
    \item {\tt uniqueID} - Unique ID of the source which is fairly static and based on {\tt xmax} and {\tt ymax}.
    \item {\tt xcen} and {\tt ycen} - Flux weighted position of the source in image coordinates.
    \item {\tt xmax} and {\tt ymax} - Flux maximum position of the source in image coordinates.
    \item {\tt RAcen} and {\tt Deccen} - Flux weighted position of the source in sky coordinates.
    \item {\tt RAmax} and {\tt Decmax} - Flux maximum position of the source in sky coordinates.
    \item {\tt sep} - Radial offset between central and maximum flux positions in units of arcsec.
    \item {\tt S\_peak} - The peak flux density in the brightest pixel of the source in units of mJy/beam calculated by multiplying {\tt cenfrac}, {\tt S\_int} and the beamsize (13.50).
    \item {\tt S\_int} - The integrated flux density of the source in units of mJy.
    \item {\tt cenfrac} - The fraction of flux in the brightest pixel.
    \item {\tt N50, N90, N100} - The number of brightest pixels containing 50, 90 and 100\% of the flux respectively.
    \item {\tt R50, R90, R100} - The approximate elliptical semi-major axis containing 50, 90 and 100\% of the flux respectively in units of arcsec.
    \item {\tt con} - The concentration of flux given by {\tt R50/R90}.
    \item {\tt semimaj} and {\tt semimin} - The weighted standard deviations along the major and minor axes i.e. the semi-major/minor first moment in units of pixels.
    \item {\tt axrat} - The axial ratio as given by {\tt semimin/semimax}.
    \item {\tt ang} - The orientation of the semi-major axis in degrees with the convention that 0 is vertical and rotates anticlockwise.
    \item {\tt SNR} - The SNR of the source calculated by\\ {\tt S\_peak/sky\_RMS\_mean} used to perform the $5\sigma$ cuts.
    \item {\tt signif} - The approximate significance of the detection using the Chi-Square distribution.
    \item {\tt S\_int\_toterr} - The total error on the flux density of the source including flux calibration errors.
    \item {\tt S\_int\_err} - The total measurement error on the flux density of the source.
    \item {\tt S\_int\_err\_sky} - The sky subtraction component of the total error on the flux density of the source.
    \item {\tt S\_int\_err\_skyRMS} - The sky RMS component of the total error on the flux density of the source.
    \item {\tt S\_int\_err\_cal} - The flux calibration component of the total error on the flux density of the source.
    \item {\tt sky\_mean} - The mean flux of the sky over all segment pixels.
    \item {\tt sky\_sum} - The total flux of the sky over all segment pixels.
    \item {\tt sky\_RMS\_mean} - The mean value of the sky RMS over all segment pixels.
    \item {\tt iter} - The iteration number when the source was flagged as having convergent flux.
    \item {\tt origfrac} - The ratio between the final converged flux and the initial iso-contour estimate.
    \item {\tt Norig} - The number of pixels in the original non-dilated segment and is $\geq$pixcut by construction.
    \item {\tt Separation} - The radial distance between a source and the nearest bright ({\tt S\_peak}$\geq$50\,mJy) source in arcsec if it is within 180 arcsec of a bright source.
\end{itemize}

\end{document}